\documentclass[conference,compsoc]{IEEEtran}
\IEEEoverridecommandlockouts
\usepackage{cite}
\usepackage{hyperref}
\usepackage{amsmath,amssymb,amsfonts}
\usepackage{algorithm}
\usepackage{algorithmic}
\usepackage{graphicx}
\usepackage{textcomp}
\usepackage{xcolor}
\usepackage{caption}
\usepackage{subcaption}
\usepackage{listings}
\usepackage{multicol}
\usepackage{multirow}

\definecolor{codegreen}{rgb}{0,0.6,0}
\definecolor{codegray}{rgb}{0.5,0.5,0.5}
\definecolor{codepurple}{rgb}{0.58,0,0.82}
\definecolor{backcolour}{rgb}{0.95,0.95,0.92}

\lstdefinestyle{mystyle}{
    backgroundcolor=\color{backcolour},   
    commentstyle=\color{codegreen},
    keywordstyle=\color{magenta},
    numberstyle=\tiny\color{codegray},
    stringstyle=\color{codepurple},
    basicstyle=\ttfamily\footnotesize,
    breakatwhitespace=false,         
    breaklines=true,                 
    captionpos=b,                    
    keepspaces=true,                 
    numbers=left,                    
    numbersep=5pt,                  
    showspaces=false,                
    showstringspaces=false,
    showtabs=false,                  
    tabsize=2
}

\newcommand{\linebreakand}{%
  \end{@IEEEauthorhalign}
  \hfill\mbox{}\par
  \mbox{}\hfill\begin{@IEEEauthorhalign}
}

\def\BibTeX{{\rm B\kern-.05em{\sc i\kern-.025em b}\kern-.08em
    T\kern-.1667em\lower.7ex\hbox{E}\kern-.125emX}}
\begin{document}

\title{Leaking Queries On Secure Stream Processing Systems
}


\author{
\centering
\IEEEauthorblockN{Hung Pham}
\IEEEauthorblockA{
\textit{Hanoi University of Science and Technology}\\
Hanoi, Vietnam \\
phammanhhung10501@gmail.com}

\and

\IEEEauthorblockN{Viet Vo}
\IEEEauthorblockA{
\textit{Swinburne University of Technology}\\
Victoria, Australia \\
vvo@swin.edu.au}

\and

\IEEEauthorblockN{Tien Tuan Anh Dinh
\IEEEauthorrefmark{1}}
\IEEEauthorblockA{
\textit{Deakin University}\\
Victoria, Australia \\
anh.dinh@deakin.edu.au}

\and

\IEEEauthorblockN{Duc Tran}
\IEEEauthorblockA{
\textit{Hanoi University of Science and Technology} \\
Hanoi, Vietnam \\
ductq@soict.hust.edu.vn}

\and

\IEEEauthorblockN{Shuhao Zhang}
\IEEEauthorblockA{
\textit{Huazhong University of Science and Technology}\\
Wuhan, China \\
shuhao\_zhang@hust.edu.cn}

\thanks{* Corresponding author}
}

\maketitle

\begin{abstract}
Stream processing systems are important in modern applications in which data arrive continuously and need to be processed in real time. Because of their resource and scalability requirements, many of these systems run on the cloud, which is considered untrusted. Existing works on securing databases on the cloud focus on protecting the data, and most
systems leverage trusted hardware for high performance. However, in stream processing systems, queries are as sensitive as the data because they contain the application logics.  

We demonstrate that it is practical to extract the queries from stream processing systems that use Intel SGX for securing the execution engine. The attack performed by a malicious cloud provider is based on timing side channels, and it works in two phases. In the offline phase, the attacker profiles the execution time of individual stream operators,
based on synthetic data. This phase outputs a model that identifies individual stream operators.  In the online phase, the attacker isolates the operators that make up the query, monitors its execution, and recovers the operators using the
model in the previous phase. We implement the attack based on popular data stream benchmarks using SecureStream and NEXMark, and demonstrate attack success rates of up to 92\%. We further discuss approaches that can harden streaming processing systems against our attacks without incurring high overhead.
 
\end{abstract}

\begin{IEEEkeywords}
side channel attack, sgx, streaming
\end{IEEEkeywords}

\section{Introduction}
\label{sec:intro}
Stream processing systems process an infinite amount of data in a continuous manner.  They are essential components in
many modern applications. For example, Internet of Things (IoT) applications that process continuous streams of data
generated by large numbers of devices~\cite{sasaki2021survey}. Other examples include e-commerce websites~\cite{tucker2008nexmark}, social
networks~\cite{noghabi2017samza, toshniwal2014storm}, and network monitoring systems~\cite{bhatia2021mstream}, which process large volumes of data generated at high
speed.

A query in a stream processing system is represented by a directed acyclic graph (DAG) of {\em operators}, for example, Map, Filter, and Join operators. Data flows through the DAG, and each operator can maintain some states during query
execution. State-of-the-art stream processing systems partition the DAG over multiple machines, such that operators can
be executed in parallel. Examples of such distributed stream processing systems (DSP) include platforms such as Flink \cite{carbone2015apache},
Storm \cite{toshniwal2014storm}, and Spark Streaming \cite{zaharia2013discretized}. These systems are able to scale up to handle large and high-velocity streams while ensuring correctness and fault tolerance~\cite{lin2016streamscope}. Most of them run on a cloud platform to leverage the latter's resources and
reliability. There are data streaming services, which are built on top of DSP, offered by major cloud
providers, for example, AWS's Kinesis and Azure's Stream Analytics.

Securing DPS against untrusted cloud providers requires protecting both the data {\em and the queries}. In general, the data stream may contain sensitive information, such as physiological readings, locations, or network activities. 
However, in many applications, the query executed over the stream can even be more sensitive and critical when they represent proprietary algorithms for data processing. One example is Bargain Index \cite{spade}, which finds bargains in the stock trading market. The query ingests a trading stream and a quote stream (sell orders), then computes bargain indexes. This query enables financial institutions to make trading decisions, thus it is an important business asset. An attacker (e.g. a competitor) knowing the query can steal and even optimize it to make faster, better trading decisions to gain profits at the expense of the victim. Another example is Network Spike Detection \cite{SAQL}, which tracks network usage and raises alerts when it exceeds a predefined threshold, indicating potential data exfiltration. An attacker knowing the query, especially the predefined threshold, can adapt its behavior to avoid detection. 


Existing works on protecting outsourced database systems, of which DPS is an instance, often leverage homomorphic encryption, secure multiparty computation, 
trusted hardware, or a combination thereof \cite{schuster2015vc3, popa2011cryptdb,reijsbergen2023tap,dinh2015m2r,zhang2015integridb}.
Given the stringent latency requirement
in stream processing, state-of-the-art secure DPS systems typically use trusted hardware  \cite{DBLP:journals/corr/abs-1805-01752,
234964,schuster2015vc3, dauterman2022waldo}. These systems run the stream engine inside trusted execution environments (TEEs), e.g., Intel SGX enclaves. 
The data is encrypted and is only processed in plaintext inside the TEEs.

In this study, we show that DSP systems that run vanilla stream engine inside the trusted hardware are vulnerable to
timing channel attacks. Specifically, we consider systems similar to Flink \cite{carbone2015apache}, which run the query DAG inside the 
Intel SGX enclaves. Our attack leverages two key insights. 
First, because of the distributed design, it is possible to
isolate individual operators from the query DAG and to run them in separate enclaves. 
By monitoring the enclave
execution, the attacker can extract fine-grained timing information of each operator. Second, the attacker can run their
own enclaves that execute {\em similar} operators with {\em similar} data to those of the target query. This is because
the data schema can be known by the attacker, for example, the attacker being a valid user of the application.
Furthermore, most DSP software is modular and open source, allowing the attacker to create enclaves for operators of
their choice and execute them with their own synthetic data.

\begin{figure}
    \centering
    \includegraphics[width=0.49\textwidth]{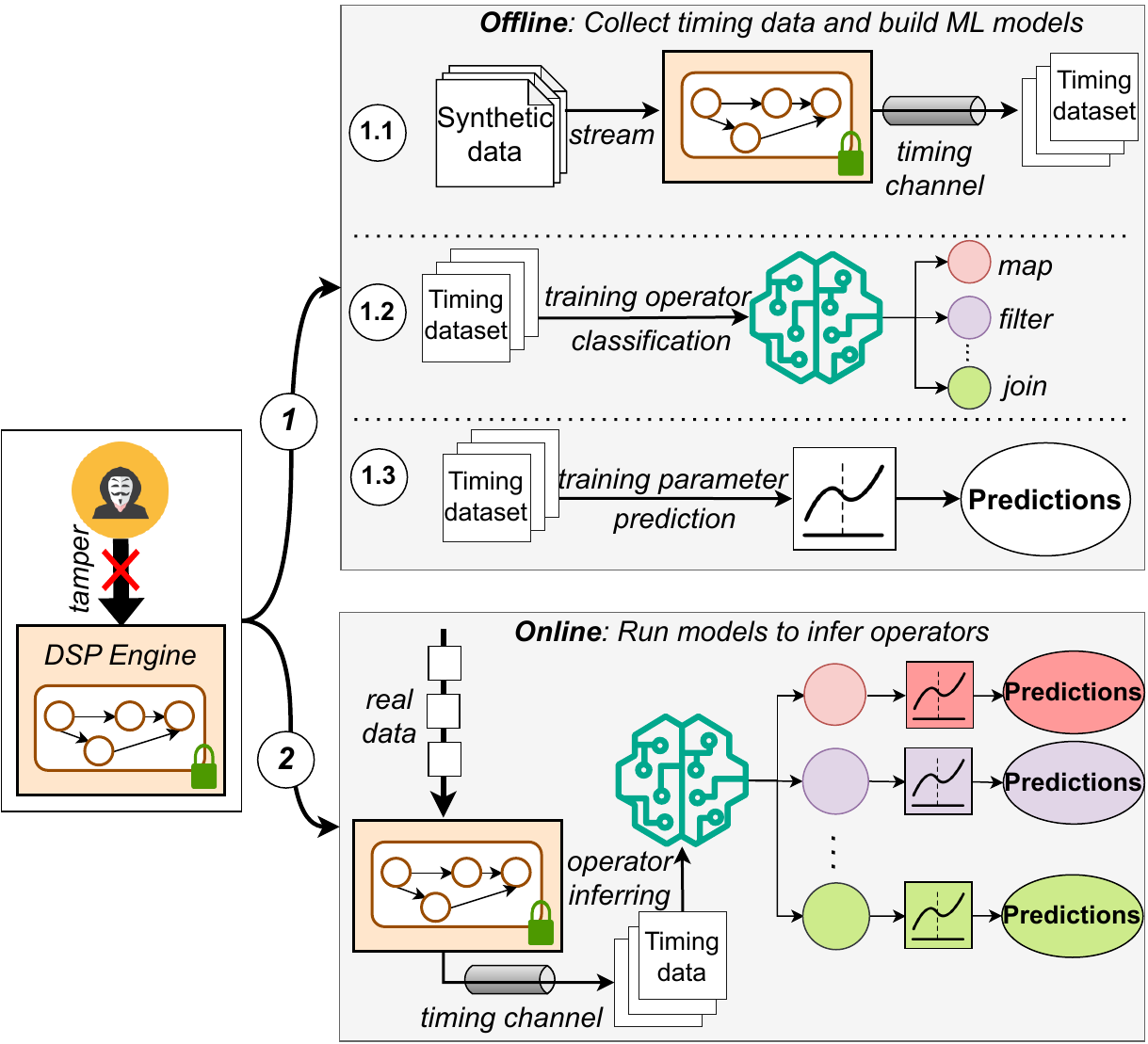}
    \caption{Overview of the attack, with an offline phase and an online phase.}
    \label{fig:attack-flow}
\end{figure}

Figure~\ref{fig:attack-flow} illustrates how the attacker exploits the two insights above to leak the content of the
query DAG. The attack occurs in two phases. In the offline phase, the attacker profiles the timing characteristics of
individual operators. 
In particular, for each operator, it creates an enclave and executes it with synthetic data while
recording the execution time. The result is a large, labeled dataset of execution times. At the end of this phase, the
attacker trains a machine learning model that predicts the operator based on the execution times. In the online phase,
the attacker isolates individual operators from the target query, such that each operator is executed in a separate
enclave. This is possible because the attacker has control over resources that the DSP system assigns to the query. It
monitors each enclave's execution and uses the model in the offline phase to learn the content and the parameters of the
operator.

We implement our attack against a customized DSP execution engine. We use popular streaming benchmarks to generate
synthetic data and queries for the offline phase. We test the trained models against unknown queries from the same
benchmarks. The results demonstrate that even with simple models, the operators can be leaked with high accuracy, of up
to $89\%$. The models can also recover the parameters inside each operator with high accuracy. The main cost
of the attack is the offline phase, but it can be significantly reduced because this phase is highly parallelizable. Our
results suggest that timing channels are a viable attack vector against DSP systems. We discuss system-level,
lightweight mitigations that undermine the offline phase by batch scheduling or operator merging.

In summary, we make the following contributions:
\begin{itemize}
\item We propose a side channel attack on distributed stream processing systems that runs a vanilla execution engine
inside Intel SGX. The attack is based on profiling the execution time of individual query operators.
\item We demonstrate that the attack can leak query operators with high accuracy, using popular streaming benchmarks.
The artifact is published at \url{https://github.com/PhamHung2020/Simple-SGX-based-DSP-Engine}.
\item We discuss lightweight mitigation approaches to harden against attacks by avoiding the collection of precise timing information, thereby making it more difficult to leak query operators.
\end{itemize}

The remainder of this paper is organized as follows. Section \ref{sec:background} provides the background on distributed
stream processing systems and Intel SGX. Section~\ref{sec:model} follows with the system and threat model.
Section~\ref{sec:methodology} details the timing side-channel attack. Section~\ref{sec:experiments} evaluates the
attack's success rate. Section~\ref{sec:discussion} discusses mitigation approaches and other limitations of the attack.
Section~\ref{sec:related} discusses the related work before Section~\ref{sec:conclusions} concludes.

\section{Background}
\label{sec:background}

\subsection{Data Stream Processing}
\label{sec:data_stream_processing}

\begin{figure}
    \centering
    \includegraphics[width=0.49\textwidth]{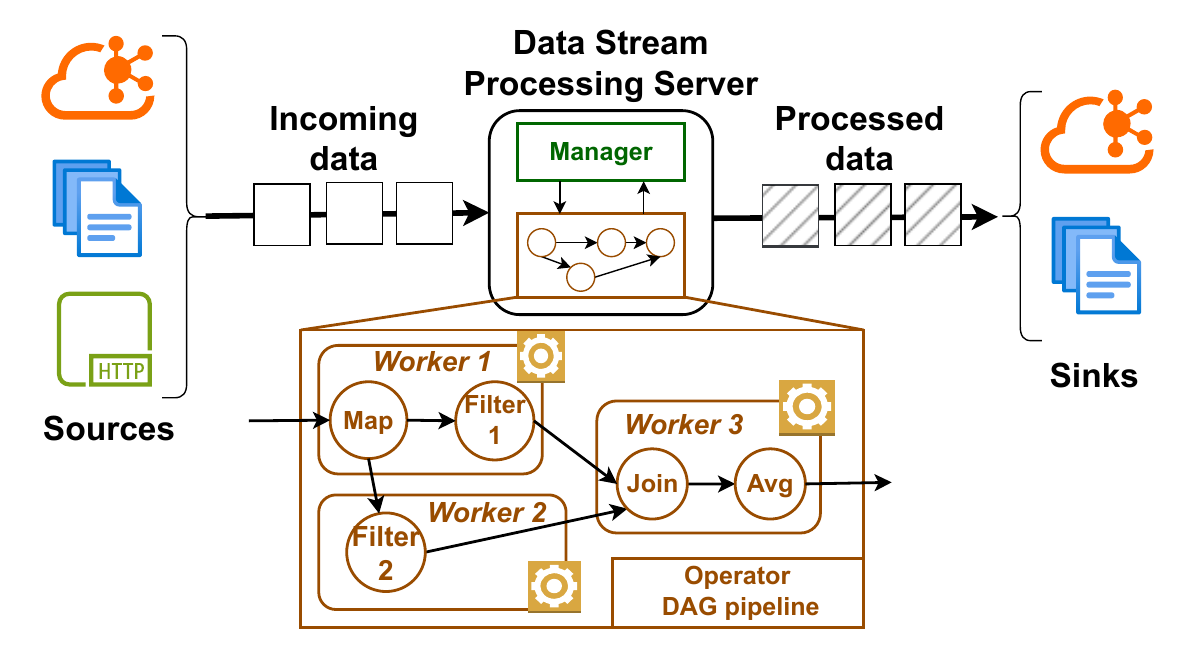}
    \caption{Data stream processing model.}
    \label{fig:data-stream-illustration}
\end{figure}

A data stream is a set of \textit{data events} (or data records) that is produced incrementally over time, rather than being fully available before the processing starts, as illustrated in Figure \ref{fig:data-stream-illustration}. Data streams typically consist of real-time data that might be unbounded. As a result, data stream processing systems are unable to control the rate at which data events arrive or store the entire stream in an accessible manner. In contrast to traditional data-management infrastructure, streaming systems have to process data events in real time using constrained memory. Continually arriving data events usually have at least one timestamp, which can be assigned to the events at the data source or on arrival.

The \textit{stream pipeline} is used by stream processing systems (or engines) to process incoming data. As data events arrive at the servers, the stream engine ingests these events, pushes them through the pipeline to perform necessary computations, and produces results at the end of the pipeline. A stream pipeline is a group of stream \textit{operators}, e.g \textit{Map} and \textit{Filter}, which are connected to each other to form a directed acyclic graph (DAG). Each operator applies a type of transformation or calculation on a single event or a limited number of events (called a \textit{window}) over the stream. The output of an operator becomes the input of the next operator in the DAG pipeline until the last operator. Figure \ref{fig:data-stream-illustration} presents an example of DAG pipelines with 4 operators: \textit{Map, Filter, Join,} and \textit{Join}. Data events are first ingested by Map, and then the outputs of Map are fed into two Filter operators. Results produced by two Filter operators are pushed into the Join and Average operators to produce final results.


Some operators need to window the data stream before processing. Windowing slices up the data stream into finite chunks for processing as a group. When working with unbounded data, windowing is required for some operators (aggregations, join, etc.) but unnecessary for others (map, filter). Three main types of windows in stream processing are (1) {\em tumbling window} in which there is no overlap between two adjacent windows, (2) {\em sliding window} in which adjacent windows overlaps, and (3) {\em session window} whose size is defined via timeouts. The window size and sliding step (in the case of sliding windows) can be event-based, i.e., adding or removing events from the window based on the number of events, or time-based, which means the timestamp of events inside the window must fit into a period of time starting from a specific time point.

\subsection{Trusted Execution Environment and Intel SGX}
\label{sec:trusted_execution_environment}

A trusted execution environment (TEE) is a secure area of a main processor. It helps the code and data loaded inside it be protected with respect to confidentiality and integrity. Data confidentiality prevents unauthorized entities from outside the TEE from reading data, while code integrity prevents code in the TEE from being replaced or modified by unauthorized entities, which may also be the computer owner itself. This is done by implementing unique, immutable, and confidential architectural security, which offers hardware-based memory encryption that isolates specific application code and data in memory. This allows user-level code to allocate private regions of memory, called \textbf{\textit{enclaves}}, which are designed to be protected from processes running at higher privilege levels.

TEEs are supported in several hardware technologies in modern processors successively, including Intel SGX \cite{SGX}, ARM TrustZone \cite{armtrustzone}, AMD's SEV. This work focuses on Intel SGX, which guarantees that an attacker with physical access to a machine will not be able to tamper with the application data without being noticed. The CPU package represents the security boundary. Data belonging to an enclave is automatically encrypted and authenticated when stored in main memory. A memory dump on a victim’s machine will produce encrypted data. A remote attestation protocol allows a user to verify that an enclave runs on a genuine Intel® processor with SGX. 



We use the first version of SGX, in which the enclave's memory, or enclave page cache (EPC), is a 128 MB area of memory predefined at boot to store enclaved code and data. At most around 90 MB can be used by the application’s memory pages, while the remaining area is used to maintain SGX metadata. Any access to an enclave page that does not reside in the EPC triggers a page fault. The SGX driver interacts with the CPU to choose which pages to evict. The traffic between the CPU and the system memory is kept confidential by the memory encryption engine (MEE). If a cache miss hits a protected region, the MEE encrypts or decrypts data before sending to or fetching from, respectively, the system memory and performs integrity checks. Data can also be persisted on stable storage protected by a seal key. 

\section{System and Threat Model}
\label{sec:model}

\subsection{System Model}
\label{sec:model-system}
There are three entities in our system model: an application owner, a cloud provider, and the users. The application
owner runs an application, for example, IoT or an e-commerce, to which multiple users send their data. The owner deploys
a distributed stream processing (DSP) system on the platform managed by the cloud provider. The owner builds a stream query
that is to be executed on the deployed DSP. The cloud provider provisions and manages the resources, and we assume that
the cloud hardware contains Intel SGX. 

\begin{figure}
    \centering
    \includegraphics[width=0.48\textwidth]{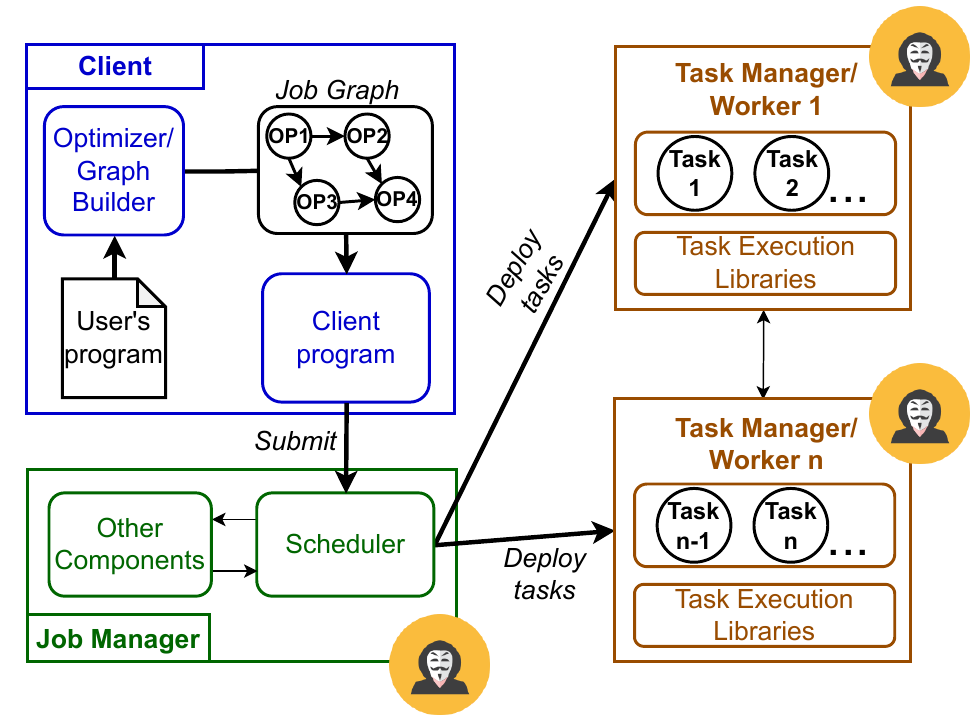}
    \caption{Architecture of the distributed stream processing system.}
    \label{fig:flink-architecture}
\end{figure}

Figure~\ref{fig:flink-architecture} illustrates the typical architecture of DSP systems, which is adopted by Apache Flink~\cite{carbone2015apache}. 
The Flink system can process  unbounded data streams with low latency, high throughput, and support
stateful computations. 
It uses a master-worker model, with a central Job Manager that coordinates query execution, and
several Task Managers that perform the actual computation and forward the intermediate results among themselves. Given a
query, the owner invokes the Graph Builder that translates it to the query DAG which is represented by the Job Graph.
This component is an optimized, lower-level representation of the query DAG, in which multiple operators might be "chained"
together into a single Task for efficiency. The owner submits the Job Graph to the Job Manager, which typically includes
the application file (e.g., Jar file in Flink) and any other dependencies.  The Job Manager invokes the Scheduler, which allocates Tasks to the available Task Managers. 

A Task Manager executes the Tasks. There is typically one Task Manager per physical machine (or node), since the Task Manager consists of multiple {\em slots}, each of which can execute one Task. The Job Manager dispatches Tasks to these slots.  It communicates with the runtime components in the Task Managers, e.g., the Actor system in Flink, to specify the resource requirements and how the intermediate data is exchanged among the Tasks. The data from the user
are sent to the first Task, that is, the source node, and traverse through the DAG of Tasks. 

We assume that the Graph Builder is run by the application owner, whereas the Job Managers and Task Managers are managed by the cloud. Specifically, the latter two run inside SGX enclaves, and all the communications to and from the enclaves are protected with SSL. Data sent by the user is encrypted such that it can only be decrypted inside the enclaves belonging to the applications. Specifically, the user authenticates with the application through a gateway, which also runs inside an enclave. This gateway establishes secure channels with the corresponding Job Manager and Task Managers after performing attestation. The user sends the data through a secure channel to the gateway, which then forwards the data to the application's source Task through another secure channel.  

Our attack focuses oonly on the execution at individual Task slots. Therefore, we implement a simplified Task Manager that has a single Task slot. Our implementation retains the same execution model as Flink's, that is, data flows
through a series of Tasks, each processing one data record at a time. We leave the implementation of Job Manager nd other production features necessary for performance and scalability to future work.


\subsection{Threat Model}
\label{sec:model-threat}

\textbf{Attacker's goals}. 
In our system model, when deploying secure streaming jobs on a TEE-enabled cloud, such as Azure Confidential Computing using the Apache Flink framework, we consider the cloud provider as the attacker. This can be due to the cloud being malicious, or being compromised by external attackers~\cite{Thalesgroup22,AkamaiIntel20,Upguard19,Bitdefender20,Zdnet19}.
%
%
However, unlike conventional adversarial goals of data breaches, revealing the confidentiality of streaming data as in previous works~\cite{DBLP:journals/corr/abs-1805-01752,
234964,schuster2015vc3, dauterman2022waldo}, the attacker is interested in compromising the privacy of query operations.
Specifically, this privacy information reveals which operations, i.e., Filter, Map, Reduce, and Join, are being executed inside the enclaves.
We stress that query operations are sensitive and critical because they contain proprietary application logic of the end-user.

\subsubsection*{Attacker's capabilities} 
{We assume the attacker has full control over the
hardware and software stack on the cloud, in line with previous studies~\cite{Silva19, sasy2018zerotrace, xiao2021privacy, Sasy19, Kim17, Ahmad18}
}. 
%
%
Specifically, the attacker can intercept all the enclave's interactions with the external environments, including the OS and host process. It can interrupt the enclave execution and provide arbitrary input to the ecalls and return its own data for the ocalls. 
The attacker can roll back the
states outside of the enclave~\cite{matetic2017rote,strackx2016ariadne}. We assume that the code inside the enclave, 
including the application code and the enclave runtime, are free of bugs. Furthermore, we assume Intel SGX hardware is secure and
it is free from micro-architectural side-channel attacks~\cite{van2018foreshadow, murdock2020plundervolt,van2020lvi}. 
The attacker targets the Job Manager and Task Manager components of the DSP, as shown in Figure~\ref{fig:flink-architecture}. In particular, the application query and its corresponding DAG, i.e., the output of the Job Graph component, are outside the cloud. The application owner submits the Job Graph to the Job Manager via a secure channel, which is not accessible by the attacker.

\subsubsection*{Attacker's knowledge} The attacker knows the source code of the DSP system deployed on the cloud, as the system is open source. Specifically, it knows how the components in Figure~\ref{fig:flink-architecture} work in their default settings. We assume that the attacker knows the data schema, that is, the fields in the data records. This knowledge can be obtained via the information about the data sources, for example, by knowing the specification of the IoT device, the attacker can have samples of the data generated by the device. Alternatively, the attacker can register as a valid user of the target application and, therefore, know what data is sent to the application. 

\subsubsection*{Knowledge of the query DAG} We assume that the attacker knows the DAG structure, but not the content of individual operators. Specifically, it knows how many operators there are in the DAG, and the data flow among them, but it does not know the operators' content. In the example in Figure~\ref{fig:data-stream-illustration}, it does not know that the first operator is a map and the two operators after that are filters. There are two scenarios in which the attacker gains knowledge of the DAG. First, the application owner does not encrypt the DAG structure. This happens when the application owner trusts the cloud for scheduling the DAG execution, therefore allowing the Scheduler component (in Figure~\ref{fig:flink-architecture}) to run outside the enclave. In this scenario, the Job Graph output contains the encrypted operators and the data flow between them, which are sent to the Job Manager via non-secure channels. Second, even when the entire DAG is encrypted, the attacker can influence the scheduler to execute one operator at a time. {For example, suppose a query has three operators, each requiring 8 GB of memory. The attacker can restrict the Task Manager’s memory to 10 GB (e.g., using \texttt {cgroups}), allowing one operator to run at a time. By monitoring which enclaves are running, the attacker learns the DAG structure.}


\subsubsection*{Out-of-scope attacks} {Unlike previous side-channel attacks~\cite{van2017telling, wang2017leaky,van2017sgx,Jaehyuk17,chen2021voltpillager}, we do not assume page fault and timer
interrupt side channels~\cite{van2017telling, wang2017leaky}. We do not consider attacks based on time-stepping the 
enclave execution~\cite{van2017sgx}, implementation bugs against TEE~\cite{Jaehyuk17}, or employing power outage attacks~\cite{chen2021voltpillager}.
These out-of-scope attacks are either orthogonal, as there have been hardware
fixes~\cite{constable2023aex}, or they are complex to execute.
For instance, MI6~\cite{Bourgeat19} and KeyStone~\cite{Lee20} can be deployed to address the attack's execution. 
Additionally, hardware providers have the option to release microcode patches to rectify these vulnerabilities~\cite{Bourgeat19,Costan161, constable2023aex}.
}

%
%
%



\section{Methodology}
\label{sec:methodology}

\subsection{Attack Overview}
As discussed in the threat model, given a target application implementing a query on a given DSP system, the attacker
knows the data schema and the structure of the query DAG. Its goal is to recover the DAG content, that is, the
identities of the DAG operators. The attacker achieves this in two phases, as depicted in Figure~\ref{fig:attack-flow}:
the offline phase and the online phase.

\subsubsection*{Offline phase}
In the offline phase, the attacker profiles individual operators that are {\em similar} to the target operators, using
{\em synthetic data}. The goal is to learn to distinguish different operators based on their execution
time. This is based on the insight that in a vanilla implementation of the DSP system, operators have different
timing (or latency) distributions. For example, a stateless operator such as map and filter will have lower latency than
a stateful operator like join. The variance in the timing distribution will also be lower for stateless operators than
for stateful ones.

The attacker first instantiates the stream processing engine with a single operator running inside an enclave. This
environment approximates the real, target environment, even though the attacker does not know which operator is running
in the target application. As discussed in our system model (Section~\ref{sec:model-system}), the attacker knows the
source code of the DSP system and the enclave code of the Job Manager and Task Manager. It then runs the system with
one Job Manager enclave and one Task Manager enclave. It constructs a Job Graph consisting of one operator of its
choice and sends it to the Job Manager enclave. The only difference between this environment and the real one is the Job
Graph containing the specification of the operator (i.e., which operator and which parameters). In the attacker-controlled environment, the attacker can specify any operator with any parameters of their choice.  

Next, the attacker creates synthetic datasets based on the known schema and the operator being profiled. In particular,
starting with the original schema, it generates smaller schemas containing subsets of the original fields and synthesizes data records with random values (of the correct field types). The diverse datasets ensure that the attacker
can capture timing profiles of the target operator under varying conditions. The attacker sends the encrypted,
synthesized data records to the system and records fine-grained execution time at the Task Manager enclave. The result is a set of time series traces per operator, where each data point in a trace corresponds to the time it takes
for the operator to process one data record.   

The execution traces form the dataset for training a two-stage machine learning model. In the first stage, a classifier
is trained to identify the type of operator based on the timing data. In the second stage, a set of regression models is
trained to infer the parameters of each operator. We focus on predicting two common parameters: the window size and the
sliding step used in windowed operators.

\subsubsection*{Online phase} 
In the online phase, the attacker runs the target application and passively collects timing data for each target operator. Even when the target query DAG comprises multiple operators, our threat model (Section~\ref{sec:model-threat}) allows the attacker to execute each operator in isolation, i.e., in an isolated Task Manager enclave. The measurement is done in the same way as in the offline phase. Given the collected data, the attacker uses the previously trained models to perform inference. First, it uses the classifier to identify which operator is running. Next, it uses the corresponding regression model to predict the operator's parameter values. Through this process, the attacker recovers the full query
DAG, thereby gaining insight into proprietary application logic.

\subsection{Attack Implementation}
\subsubsection*{DSP system}
\begin{figure}
    \centering
    \includegraphics[width=0.49\textwidth]{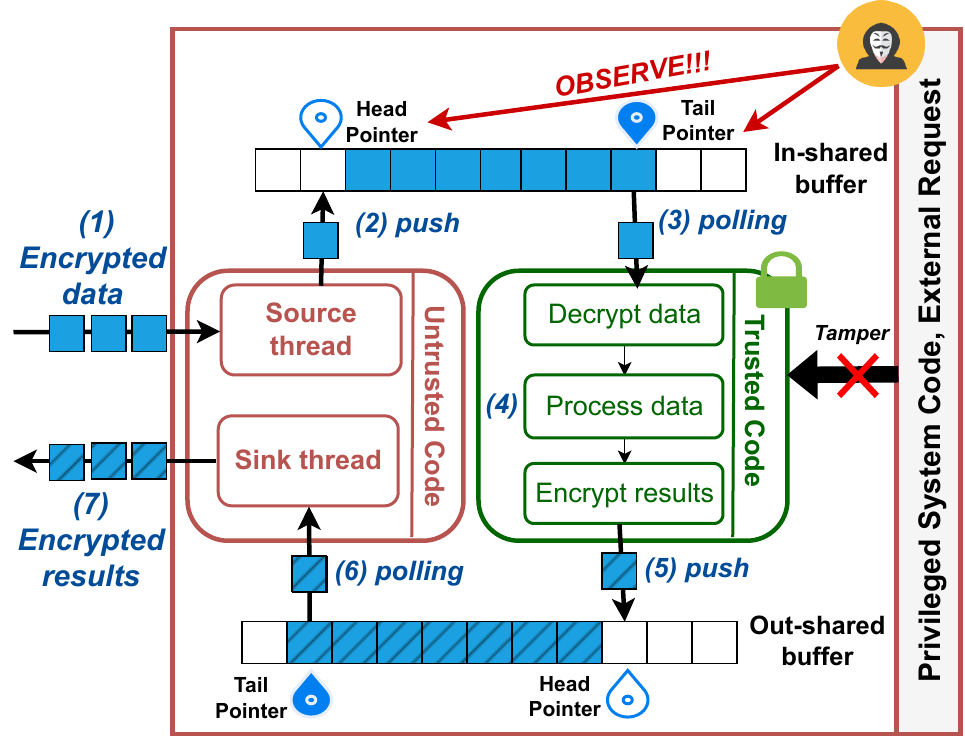}
    \caption{Overview of the DSP implementation and timing measurement against the enclave running a query operator.}
\label{fig:architecture-overview}
\end{figure}

We implement a simplified version of the DSP system to demonstrate our attack. Our implementation retains the same
execution semantics as that of a full-fledged, production-ready DSP system. The system is implemented in C++ using
Intel’s SGX SDK~\cite{sgxsdk}, and each query operator runs in isolation
within its own enclave. It uses {\em hot call}\cite{hotcall}, instead of the traditional ecall/ocall or switchless
call~\cite{switchlesscall}, as the primary I/O interface between the untrusted and trusted parts of the application. The
hot calls lead to significant performance improvement compared to the alternative I/O interfaces.  They also reduce the
enclave's exposure to known communication-based vulnerabilities. Figure~\ref{fig:architecture-overview} shows the
overview of the implementation and how the attacker measures the execution timing.

Our implementation consists of two components: an untrusted part, which manages data I/O and coordinates operator
chaining, and a trusted part running inside an enclave and executing the operator's computation logic.  It is
multi-threaded, with concurrent threads operating in a pipelined fashion. The data records are processed in five steps as follows. 

First,  the source thread, running outside the enclave, reads the encrypted data from external sources (e.g., files or network streams). It pushes the records into an in-shared buffer, which is in a memory region shared between the untrusted part and the enclave. Second, inside the enclave, an event receiver thread continuously polls the in-shared buffer for new data. When a new
data record is detected, it is pulled into the enclave using a hot call, then decrypted. Third, the decrypted data is passed to the operator, which performs the intended computation (e.g., filter, map, join, aggregate) and produces an intermediate result. Fourth, the result is sent to an event sender component within the enclave, which encrypts and pushes it into the
out-shared buffer, another shared memory region between the enclave and the untrusted part. Fifth, in the untrusted part, a sink thread (running concurrently with the source thread) continuously polls the
out-shared buffer. It fetches the encrypted results when available, then either forwards them to the next enclave in the
processing pipeline or stores them in persistent storage, such as a database or CSV file.


\subsubsection*{Shared buffers}
The shared buffers in our system are implemented as circular buffers, a data structure commonly used in multi-threaded
environments to support ordered communication between producers and consumers. A circular buffer uses a fixed-size array
in a way that treats the beginning and end as contiguous, enabling continuous data flow without moving the elements.


Each buffer maintains two pointers: a head pointer (also known as the write pointer), pointing to the position where
the next incoming data item will be written, and a tail pointer (or read pointer), pointing to the position where the
next data item will be read. When the producer writes data and the buffer is not full, the head advances to the next
slot in the array, inserting the new element. Conversely, when the consumer reads data, the element at the tail is
retrieved, and the tail pointer advances to the next unread item. If either pointer reaches the end of the array, it
wraps around to the beginning, enabling circular traversal of the buffer. This wrapping behavior makes it suitable for
streaming scenarios, especially under memory constraints. When the head and tail pointers collide, the buffer is
considered full, and further writes are blocked until space is freed.

In our implementation, each circular buffer is encapsulated as an object, exposing a \textit{push()} method used by source threads or enclaves to insert new data records, and a \textit{pop()} method used by enclaves or sink threads to retrieve available data.
The buffers enable thread-safe communication between the untrusted and trusted components of the system.

\subsubsection*{Data encryption and decryption}
All data in both in-shared and out-shared buffers is encrypted. They are decrypted only when inside the enclaves. We use the AES-GCM 128-bit scheme for encryption and decryption operations. AES-GCM  guarantees both data integrity and confidentiality. The SGX SDK provides out-of-the-box the implementation of AES-GCM, which is accessible through two functions: \textit{sgx{\_}aes{\_}gcm{\_}encrypt} and
\textit{sgx{\_}aes{\_}gcm{\_}decrypt}. In the untrusted part, we use the OpenSSL \cite{openssl} implementation of AES-GCM. For simplicity, we hardcoded a fixed key inside the enclave's source code.


\subsubsection*{Query operators}
Each operator is implemented as a function that takes one or two (e.g., for join operators) data records as input, performs
some computation, and then returns the result. The operator can be stateful, i.e., maintaining some intermediate states
between inputs. The functions operate on decrypted data: the input is decrypted inside the enclave before being passed
to the function, and the output is encrypted before being written to the out-shared buffer.
{For window operators, we implement \textit{sliding window}, in which window and sliding step sizes are based on event counts.}

\subsection{Timing Measurement}

We use the RDTSCP instruction, a variant of the RDTSC instruction, to measure the execution time of each operator in clock
cycles. On production SGX systems, RDTSC and its variants are not allowed inside the enclave, therefore we make all RDTSCP calls in the untrusted part. When using RDTSCP to measure the cycle count of a user-space operation, it 
is important to ensure that there are no context switches to the operating system, which would contaminate the
measurement. To minimize these switches, all CPU cores allocated for the enclave execution are isolated such that no
other applications are running on them.  In addition,
instead of the default scheduling algorithm of the operating system, we leverage CPU affinity to bind each thread to a
specific CPU core to minimize interference. The system runs on a data stream consisting of millions of events over a long
period of time, which helps reduce the impact of noise and other sources of timing contamination.


The attacker cannot measure processing time from inside the enclave, due to the assumption that the hardware is secure.
Therefore, all timing measurements must be done from outside the enclave, i.e., in the untrusted part. As shown in
Figure~\ref{fig:architecture-overview}, for each enclave, the measurement is performed by the Observer
thread, which continuously observes the changes in the in-shared buffer of the enclave and records the RDTSC value each
time a change is detected. Particularly, the Observer thread repeatedly checks the value of the tail pointer of the
observed buffer. If the current value of the tail pointer is different from the last recorded value, that means the
enclave has ingested a new data event for processing.  The Observer thread will record RDTSC at that moment immediately,
and the tail pointer's current value will be saved as the most recent value. Finally, the processing time for a data
record is computed by subtracting the previous recorded RDTSC value from the newly recorded one. By experimenting with
each operator with a large number of data records, the attacker can obtain a large number of processing times, which form the
dataset for training machine learning models that predict the operator logics and parameters.

\subsection{Machine Learning Model Settings}
We perform a number of pre-processing steps on the data collected from running the operators with synthetic data. These steps prepare the data for training the machine learning models. Specifically, for each time series, we discard the first and last 5\% of data points.  We remove these segments because they are noisy due to the system warming up and termination.  The remaining 90\% of the time series, capturing the steady-state data, is then transformed into fixed-length representations. 
This is necessary because the time series have varying lengths and therefore cannot be directly input into standard classifiers. We employ two transformation methods. The first is a statistical method based on the \textit{cumulative distribution function (CDF)} of the time series. The second is a representation learning method using \textit{TS2Vec}~\cite{yue2022ts2vec}, a self-supervised embedding model that generates compact representations of the time series. 

{\bf CDF-based representation.} The reason for using CDFs is that they capture the empirical distribution of data points in a compact representation,  while abstracting away the temporal order in the time series. Given a univariate time series containing operator execution times, we generate its CDF and uniformly sample $k$ data points from the result. In our implementation, we empirically set $k=1024$, obtaining a fixed-size vector of 1024 values representing the shape of the distribution.

{\bf TS2Vec-based representation.} This method transforms a variable-length time series into a fixed-length embedding vector. We use the TS2Vec model~\cite{yue2022ts2vec} that is specifically designed to capture rich temporal features and dynamics in time series data. TS2Vec works by training a self-supervised model. Due to limited resources, we trained the TS2Vec model for only 10 epochs, with the output embedding dimension set to 1024 --- the same dimensionality in the CDF-based method to enable a fair comparison. Specifically, we trained TS2Vec on the time series collected during the offline phase, and then used the learned encoder to turn each time series into an embedding vector for downstream classification and regression tasks.

{\bf Models.} We select three simple, lightweight machine learning models, namely Random Forest, Support Vector Machine (SVM), and XGBoost. These models demonstrate accuracy while remaining efficient and are amenable to model interpretability.  

\section{Evaluation}
\label{sec:experiments}
\subsection{Experiment Setup}

We demonstrate our attack on two stream query benchmarks:  SecureStream \cite{DBLP:journals/corr/abs-1805-01752} and NEXMark \cite{nexmark}. Each benchmark consists of a set of representative queries, which we deployed on our system to collect processing time data for operator classification and parameter inference.

\noindent \textbf{SecureStream.} This benchmark consists of a simple map-filter-reduce pipeline designed to process a real-world dataset released by the U.S. Bureau of Transportation Statistics. The dataset contains over $14$ million flight records from multiple air carriers. We discuss the benchmark's operators in Appendix~\ref{app:benchmark}. 

\noindent \textbf{NEXMark.} This benchmark is widely used to evaluate stream processing systems~\cite{begoli2019one,hoffmann2019megaphone,sponge2023,Meces2022,three_step_2018}. It targets online auction applications. It consists of three types of data events: Person, Auction, and Bid, representing three typical entities in the auction context. We conduct our experiments using the first six queries, covering a diverse range of operator types and dataflow patterns. The last two queries are only a replication of previous ones and are therefore omitted without loss of generality. Since our system only implements basic operators with simple window mechanisms, each query is simplified if necessary and then represented as a Directed Acyclic Graph (DAG) of implemented stream operators, which defines the flow of data through the system. A summary of the deployed queries is presented in Appendix~\ref{app:benchmark}. 

\noindent{\bf Timing dataset.}  We executed the deployed queries using a mix of collected and synthetically generated data as input. We systematically vary the query parameters. For Map and Filter operators, we replaced the fixed mapping and filtering conditions in the original queries with multiple alternative expressions. For window-based operators, we altered the window size and sliding step values across runs. Table~\ref{tab:dataset} summarizes the results. 

\begin{table}
\caption{Timing dataset.}
\label{tab:dataset}
\begin{tabular}{lc|lc}
\hline
\multicolumn{2}{c|}{\textbf{SecureStream benchmark}}                            & \multicolumn{2}{c}{\textbf{NEXMark benchmark}}                       \\ \hline
\textbf{Operator}             & \multicolumn{1}{l|}{\textbf{No. of samples}} & \textbf{Operator}   & \multicolumn{1}{l}{\textbf{No. of samples}} \\ \hline
Map                           & 350                                             & Map                 & 1262                                           \\
Filter                        & 350                                             & Filter              & 1297                                           \\
Reduce                        & 350                                             & Join                & 1009                                           \\
\multicolumn{1}{c}{\textbf{}} & \multicolumn{1}{l|}{}                           & Max                 & 1059                                           \\
                              &                                                 & Average   & 1018                                           \\
                              &                                                 & Avg (Partition) & 1071                                           \\ \hline
\textit{Total}                & 1050                                            & \textit{Total}      & 6716                                           \\ \hline
\end{tabular}
\end{table}

\subsubsection*{Hardware specifications}
Experiments are conducted on a computer running Ubuntu 22.04 equipped with Intel(R) Core(TM) i5-9300H, 4 physical cores, and 16 GB RAM, with SGX version I.

\subsection{Results}
\label{subsec:results}

\begin{figure*}
    \centering
    \begin{subfigure}[t]{0.3\textwidth}
        \includegraphics[width=0.9\linewidth]{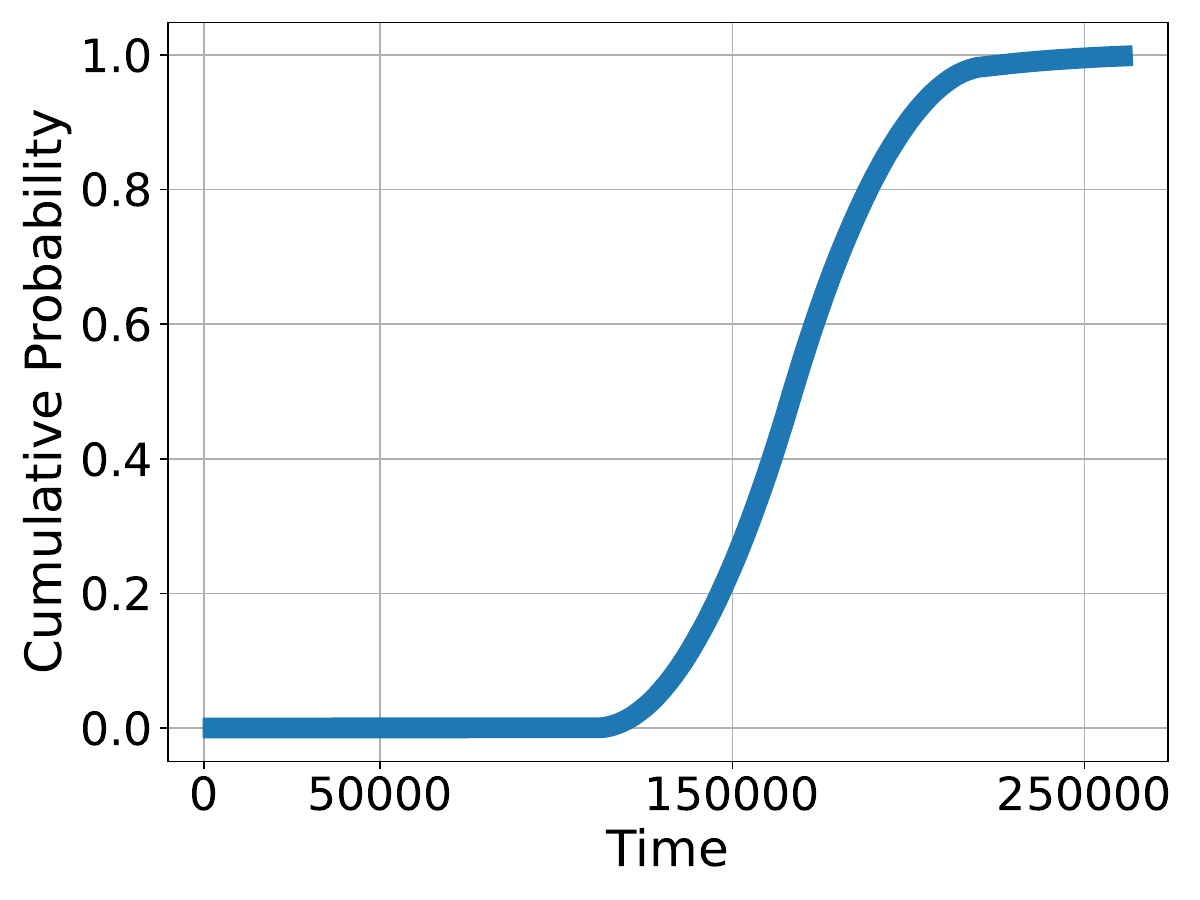}
        \caption{Map}
    \end{subfigure}%
    \hfill
    \begin{subfigure}[t]{0.3\textwidth}
        \includegraphics[width=0.9\linewidth]{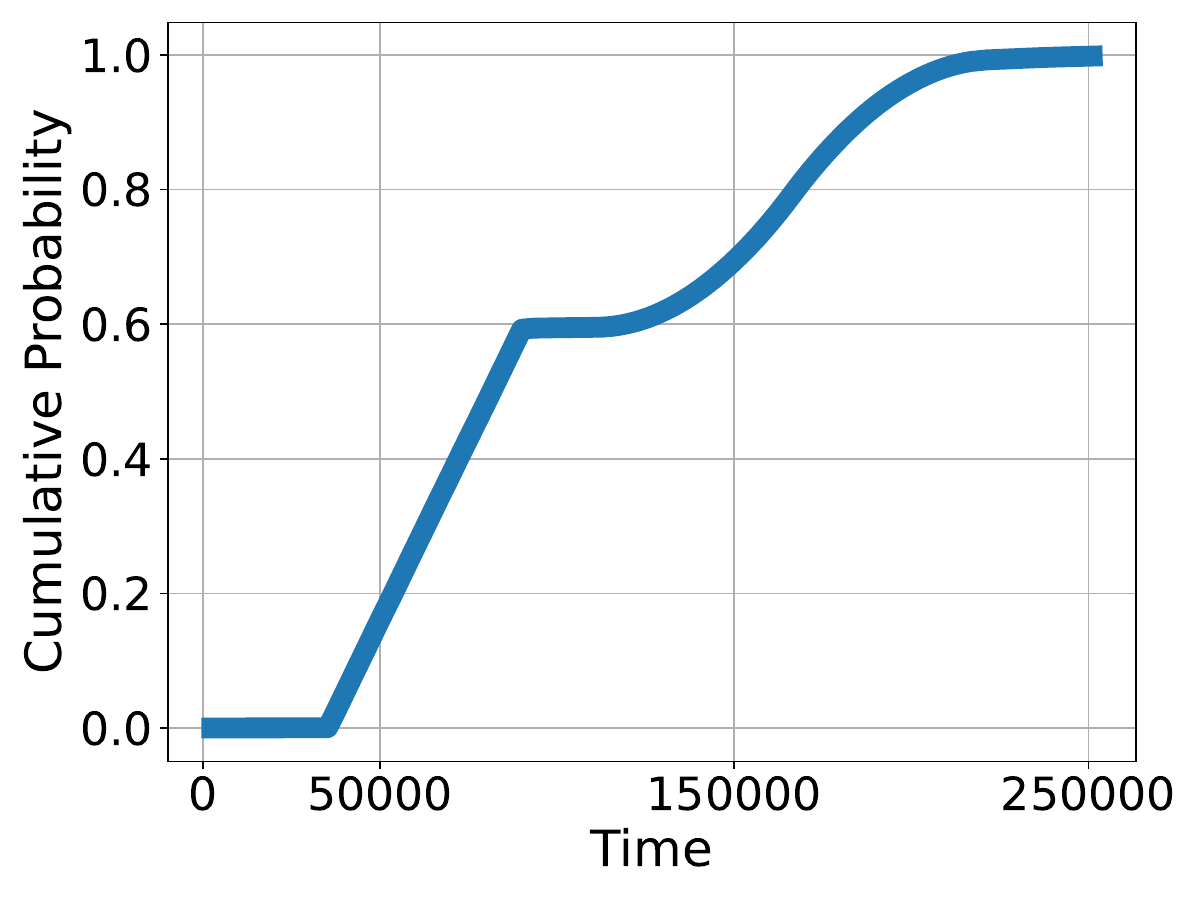}
        \caption{Filter}
    \end{subfigure}%
    \hfill
    \begin{subfigure}[t]{0.3\textwidth}
        \includegraphics[width=0.9\linewidth]{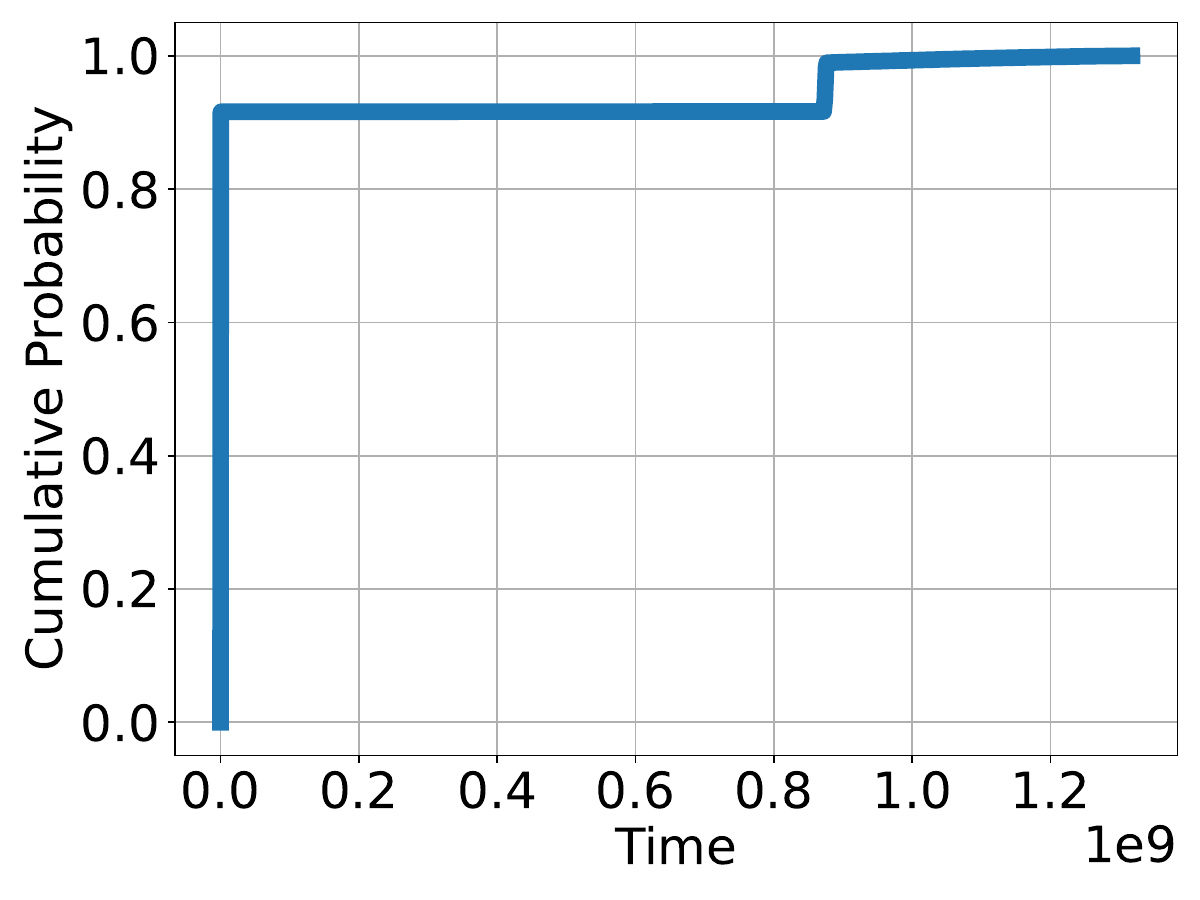}
        \caption{Join}
    \end{subfigure}%

    \vspace{1mm}
    \begin{subfigure}[t]{0.3\textwidth}
        \includegraphics[width=0.9\linewidth]{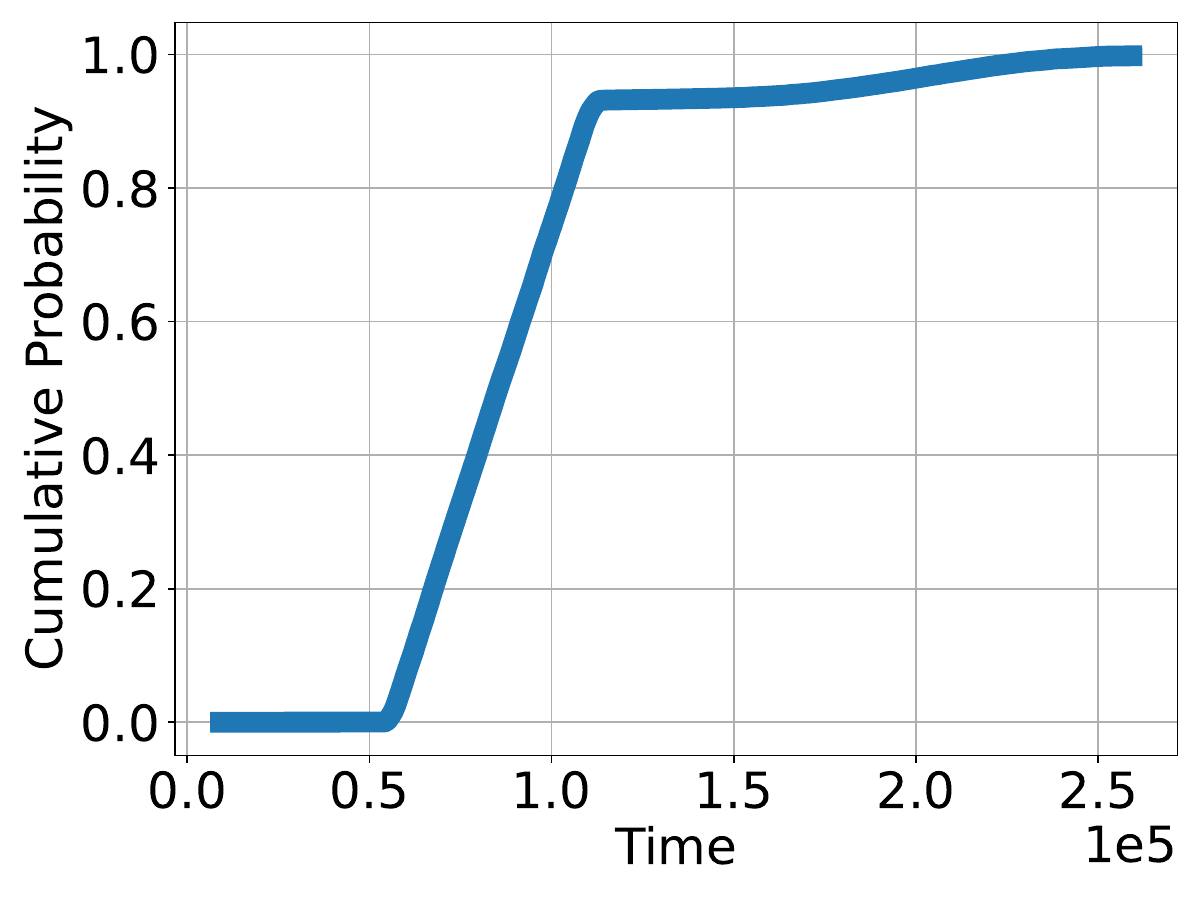}
        \caption{Average (Partition)}
    \end{subfigure}%
    \hfill
    \begin{subfigure}[t]{0.3\textwidth}
        \includegraphics[width=0.9\linewidth]{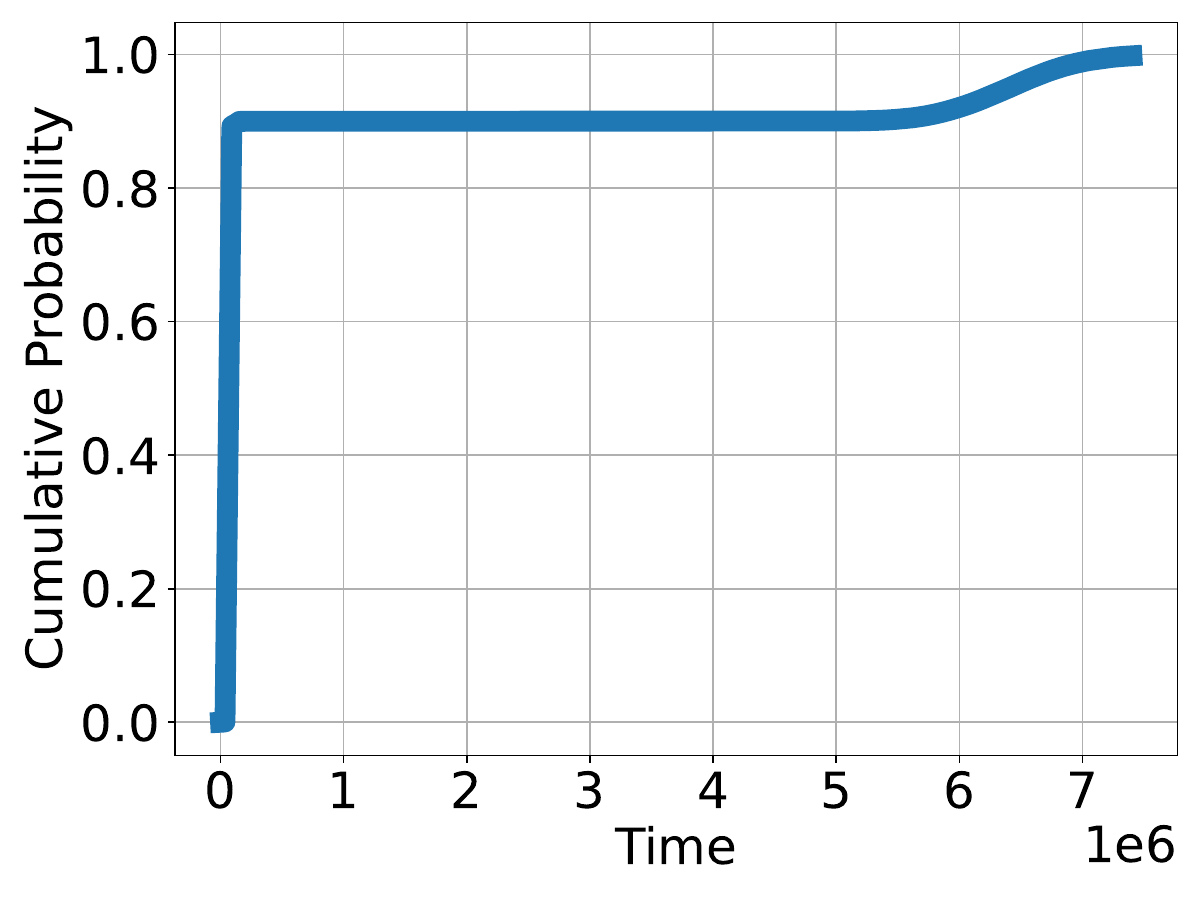}
        \caption{Count}
    \end{subfigure}%
    \hfill
    \begin{subfigure}[t]{0.3\textwidth}
        \includegraphics[width=0.9\linewidth]{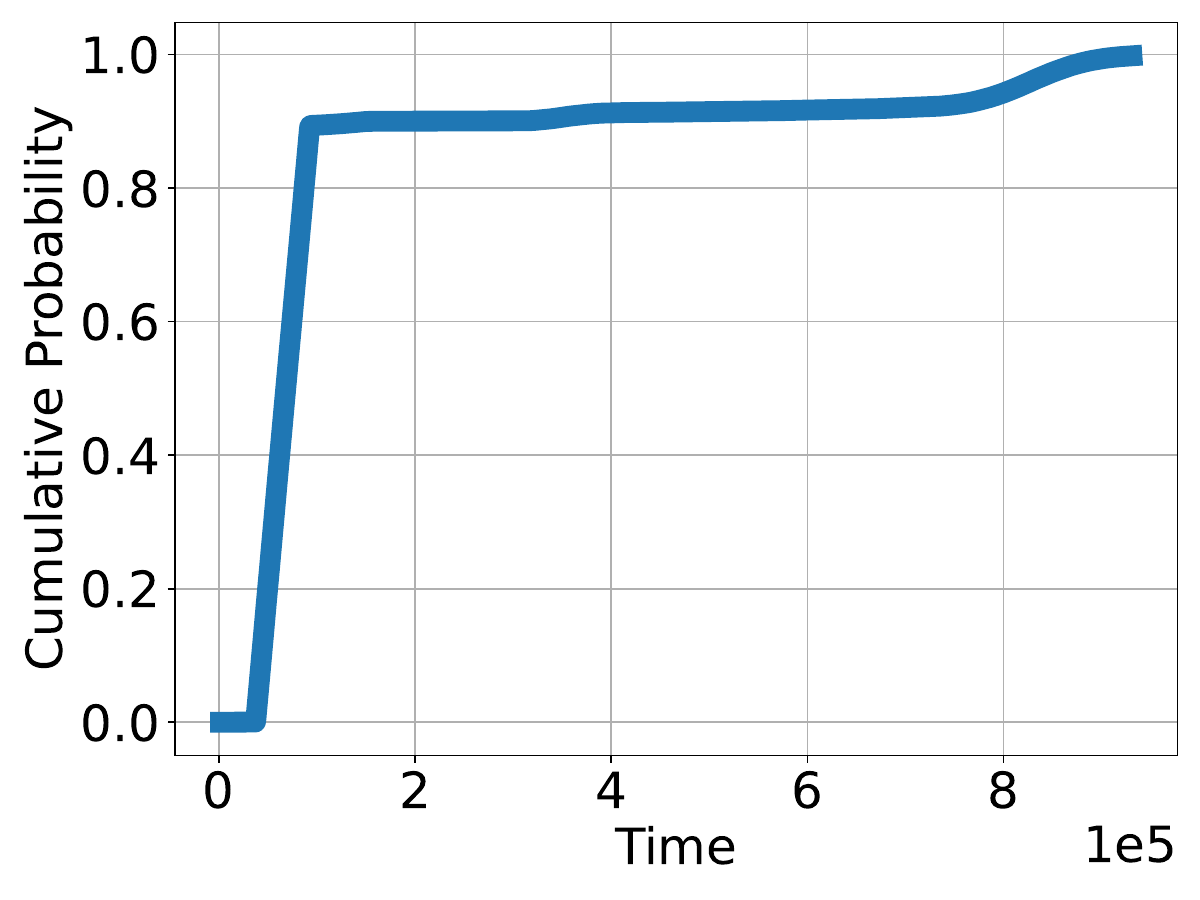}
        \caption{Max}
    \end{subfigure}%
    \caption{CDFs of operators' timing distributions.}
    \label{fig:cdf_of_each_operators}
\end{figure*}

\subsubsection{CDF-based representation} Figure~\ref{fig:cdf_of_each_operators} illustrates representative CDF curves for six operator types: Map, Filter, Join, Average, Count, and Max. The x-axis (\textit{Time}) represents the measured RDTSCP values, while the y-axis corresponds to the computed cumulative probability.

The Map operator (see Figure \ref{fig:cdf_of_each_operators}.a) exhibits a smooth, continuous CDF curve. This is expected, as Map applies a simple, stateless transformation uniformly to all incoming events, leading to consistent processing durations with minimal variance.

In contrast, the CDF for the Filter operator (see Figure \ref{fig:cdf_of_each_operators}.b) displays a noticeable “leap” or abrupt increase in the slope around the midpoint. This discontinuity reflects the operator's branching behavior: input events that satisfy the filtering condition are passed downstream, typically incurring a slightly longer processing time, while events that are filtered out are processed more quickly. This dichotomy results in two dominant processing time modes and causes the CDF to exhibit a step-like feature.

A similar pattern is observed in the Join operator (see Figure \ref{fig:cdf_of_each_operators}.c), but with some key differences. 
The leap in the Join CDF occurs at a different location along the x-axis, and the gap between the two processing modes is more pronounced. 
This is due to the nature of stream joins, which depend on the window state. When a new event arrives but the corresponding join window is incomplete or has not yet slid forward sufficiently, the operator performs only state maintenance, leading to shorter execution times. When the window becomes complete, the join computation is triggered, resulting in a significant increase in processing time—hence the sharper and wider leap in the CDF.

The final group—Average, Count, and Max—share a common structure, as they are all windowed aggregation operators (see Figures \ref{fig:cdf_of_each_operators}.d-f). 
Their CDFs exhibit similar leap patterns, akin to the Join operator, due to the windowing behavior. 
While the precise computation differs (e.g., summing for average vs. scanning for maximum), the general processing pattern remains consistent: periods of lightweight state updates punctuated by more expensive aggregation computations when the window criteria are satisfied.

These observations demonstrate that CDF curves can effectively capture structural differences in operator behavior, offering a visually intuitive and statistically meaningful representation for operator classification.

\subsubsection{TS2Vec-based representation} 
\begin{figure}
    \centering
        \includegraphics[width=0.8\linewidth]{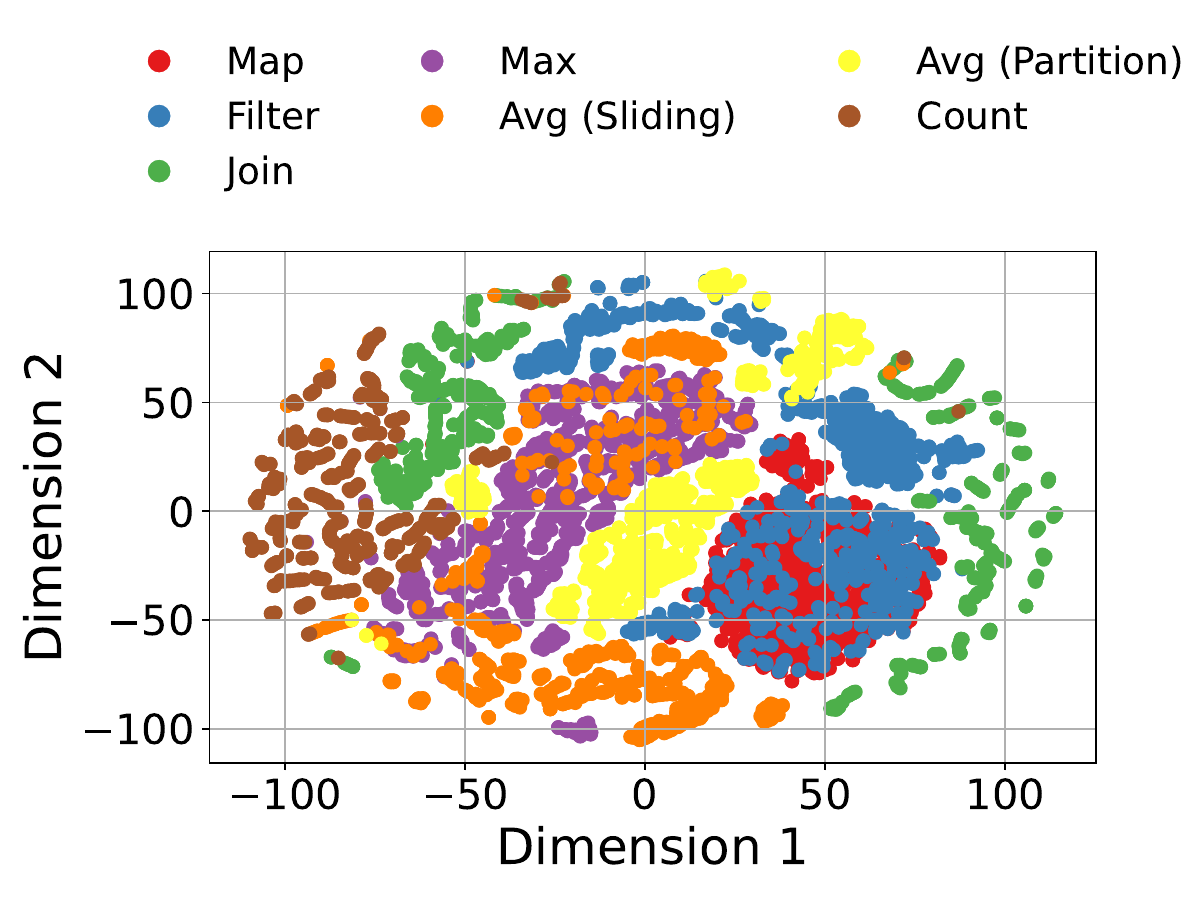}
    \caption{t-SNE visualization of NEXMark TS2Vec embeddings}
    \label{fig:ts2vec-tsne}
\end{figure}

The TS2Vec model was trained on the timing dataset for 10 epochs, with an output embedding dimension of 1024. 
Figure~\ref{fig:ts2vec-tsne} presents the resulting 2D projection of the embeddings using t-distributed Stochastic Neighbor Embedding (t-SNE). Each point corresponds to a time series representation of an operator execution, colored according to its operator type. The figure reveals that, in general, embeddings from different operators form distinguishable clusters, indicating that TS2Vec is effective in capturing operator-specific temporal patterns for classification purposes.

However, we also observe that samples from the Filter operator exhibit substantial overlap with those from the Map operator. This suggests that, in practice, distinguishing between these two stateless operators based on timing behavior alone may be challenging. The similarity in processing logic—both operators apply a single-event transformation without maintaining state—likely contributes to this ambiguity in the latent space.

\subsubsection{Operator classification on single benchmark} We evaluate the effectiveness of three widely used machine learning models—Random Forest (RF), Support Vector Machine (SVM), and XGBoost—for operator classification based on timing representations. For each model, we perform a grid search over various combinations of hyperparameters (refer to Appendix~\ref{app:evalution}), and use 5-fold cross-validation to ensure robustness and reduce overfitting. Table \ref{tab:performances_single_benchmark} reports the best classification accuracy achieved by each model under both CDF-based and TS2Vec-based representations.

\begin{table*}
\centering
\caption{Performance of operator classification models}
\label{tab:performances_single_benchmark}
\begin{tabular}{lcccccccc}
\hline
\multicolumn{1}{c}{\multirow{2}{*}{\textbf{Models}}} & \multicolumn{4}{c}{\textbf{SecureStream benchmark}}                                                                      & \multicolumn{4}{c}{\textbf{NEXMark benchmark}}                                                                           \\ \cline{2-9} 
\multicolumn{1}{c}{}                                 & \multicolumn{1}{l}{Accuracy} & \multicolumn{1}{l}{Precision} & \multicolumn{1}{l}{Recall} & \multicolumn{1}{l}{F1 score} & \multicolumn{1}{l}{Accuracy} & \multicolumn{1}{l}{Precision} & \multicolumn{1}{l}{Recall} & \multicolumn{1}{l}{F1 score} \\ \hline
\multicolumn{9}{c}{\textbf{CDF}}                                                                                                                                                                                                                                                                           \\ \hline
Random Forest                                        & 0.8503                       & 0.8399                        & 0.8316                     & 0.8311                       & 0.7178                       & 0.7757                        & 0.7440                     & 0.7262                       \\
SVM                                                  & 0.8503                       & 0.8413                        & 0.8314                     & 0.8310                       & 0.7735                       & 0.8005                        & 0.7979                     & 0.7791                       \\
XGBoost                                              & 0.8569                       & 0.8502                        & 0.8384                     & 0.8378                       & 0.6637                       & 0.6834                        & 0.7093                     & 0.6179                       \\ \hline
\multicolumn{9}{c}{\textbf{TS2Vec Embedding}}                                                                                                                                                                                                                                                              \\ \hline
Random Forest                                        & 0.8914                       & 0.9077                        & 0.8758                     & 0.8748                       & 0.8479                       & 0.8798                        & 0.8733                     & 0.8635                       \\
SVM                                                  & 0.9145                       & 0.9072                        & 0.9047                     & 0.9052                       & \textbf{0.8662}                       & \textbf{0.8916}                        & \textbf{0.8894}                     & \textbf{0.8853}                       \\
XGBoost                                              & \textbf{0.9211}                       & \textbf{0.9253}                        & \textbf{0.9102}                     & \textbf{0.9110}                       & 0.8510                       & 0.8915                        & 0.8805                     & 0.8684                       \\ \hline
\end{tabular}
\end{table*}

Overall, models trained on TS2Vec embeddings consistently outperform those trained on CDF features. On the SecureStream benchmark, all three models achieve comparable performance with CDF features, yielding approximately 85\% accuracy. When switching to TS2Vec embeddings, classification accuracy improves for all models. XGBoost achieves the highest accuracy of 92\%, representing a 7\% improvement over its CDF-based counterpart and a 1--2\% margin over the other TS2Vec-based models.

A similar trend is observed on the NEXMark benchmark. Under CDF features, none of the models surpass 80\% accuracy; notably, XGBoost performs the worst, achieving only 66\%. In contrast, using TS2Vec embeddings boosts the accuracy of all three models by at least 10\%. SVM emerges as the top-performing model on this benchmark, achieving 86\% accuracy—1--2\% higher than both Random Forest and XGBoost.

These results demonstrate that TS2Vec is a more effective feature representation for operator classification tasks, particularly in capturing the temporal structure present in timing data.


\subsubsection{Confusion matrix} The confusion matrix of the best-performing TS2Vec-based models for NEXMark is presented in Figure \ref{fig:confusion_matrix_nexmark} (we discuss both benchmarks in more detail in Appendix~\ref{app:evalution}). We observe that the model misclassifies 64\% of Map samples and 17\% of Filter samples as the other class. Additionally, the model struggles to differentiate between the Max and Average (sliding) operators. This can be attributed to their shared characteristics as aggregation functions operating over sliding windows. 

\begin{figure}
    \centering
    \begin{subfigure}[t]{0.45\textwidth}
        \includegraphics[width=\linewidth]{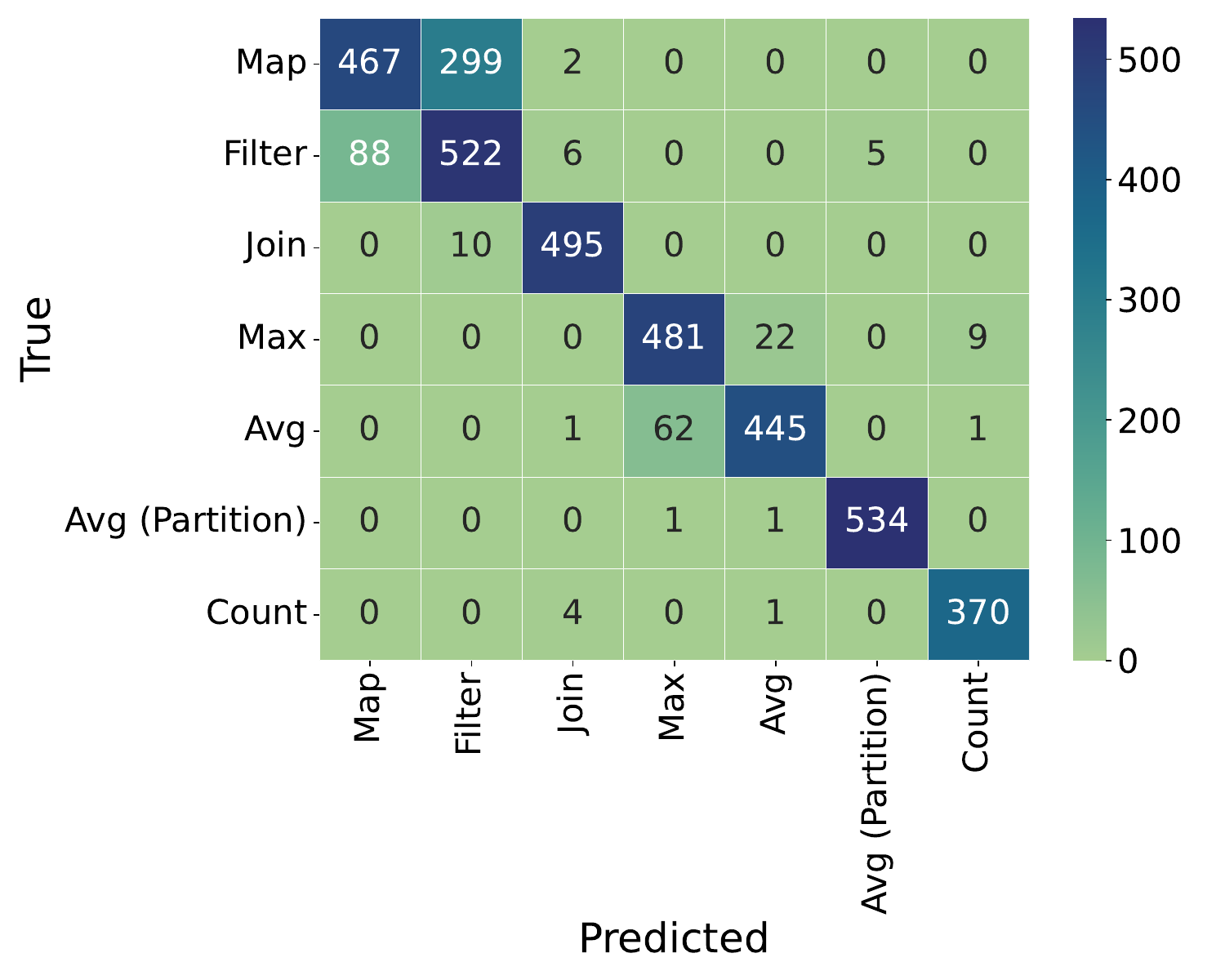}
    \end{subfigure}
    \caption{Confusion matrix (NEXMark)} 
    \label{fig:confusion_matrix_nexmark}
\end{figure}

\subsubsection{Cross-benchmark operator classification} To evaluate the generalizability of operator classification models across benchmarks, we retrained the best-performing model from the NEXMark benchmark on its entire dataset and tested its performance on the SecureStream benchmark. The \textit{Reduce} operator in SecureStream can be considered as the \textit{Average (Sliding)} operator in the NEXMark benchmark. The results, summarized in Table \ref{tab:cross_performances}, reveal a significant drop in performance when models are applied outside of their training domain.

As expected, no models achieved an accuracy above 80\% in this cross-benchmark setting. Interestingly, the best result was obtained by the CDF-based XGBoost model, which achieved 80\% accuracy, 79\% precision, 78\% recall, and a 78\% F1-score. In contrast, TS2Vec-based models, which had outperformed their CDF counterparts in within-benchmark evaluations, showed poor generalization: all TS2Vec-based models failed to surpass 50\% on any metric.

These findings indicate a trade-off between specificity and generalizability. TS2Vec embeddings are highly effective when trained and tested on the same benchmark, likely due to their capacity to learn benchmark-specific temporal patterns. On the other hand, the CDF representation, though less effective for within-benchmark classification, appears to offer greater robustness and transferability across benchmarks.

\begin{table}
\centering
\caption{Performance of NEXMark models on SecureStream benchmark}
\label{tab:cross_performances}
\begin{tabular}{lcccc}
\hline
\multicolumn{1}{c}{\multirow{2}{*}{\textbf{Models}}} & \multicolumn{4}{c}{\textbf{Metrics}}                                                                                     \\ \cline{2-5} 
\multicolumn{1}{c}{}                                 & \multicolumn{1}{l}{Accuracy} & \multicolumn{1}{l}{Precision} & \multicolumn{1}{l}{Recall} & \multicolumn{1}{l}{F1 score} \\ \hline
\multicolumn{5}{c}{\textbf{CDF}}                                                                                                                                                \\ \hline
Random Forest                                        & 0.6140                       & 0.7681                        & 0.6190                     & 0.6460                       \\
SVM                                                  & 0.7169                       & 0.7632                        & 0.7057                     & 0.7186                       \\
\textbf{XGBoost}                                              & \textbf{0.8008}                       & \textbf{0.7935}                        & \textbf{0.7842}                     & \textbf{0.7819}                       \\ \hline
\multicolumn{5}{c}{\textbf{TS2Vec Embedding}}                                                                                                                                   \\ \hline
Random Forest                                        & 0.4864                       & 0.5108                        & 0.5031                     & 0.5034                       \\
SVM                                                  & 0.4255                       & 0.4324                        & 0.4246                     & 0.4165                       \\
XGBoost                                              & 0.4790                       & 0.4751                        & 0.4751                     & 0.4751                       \\ \hline
\end{tabular}
\end{table}

\subsubsection{Query recovery}
We now evaluate the success rate of recovering the entire query, as opposed to recovering individual operators. This is an important metric because a query comprises multiple distinct operators, the attacker must identify all the operators in order to learn the query. Even when one operator is recovered with high accuracy (e.g., 92\%), it does not necessarily follow that the query can be recovered with the same level of accuracy. We use Query Recovery Successful Rate (QRSR) metric, which quantifies the degree to which a query can be reconstructed. QRSR is defined as the product of the classification accuracies of all the operators in the query.

We train and evaluate the models under two settings. In the first setting, operators from all queries are divided evenly into training and testing sets. Each model is trained on the training subset and evaluated on the testing subset. In the second setting, all operators belonging to one query are withheld for testing, while operators from the remaining queries are used for training. Experiments under this setting are repeated with a different query withheld each time. We only consider the NEXMark benchmark, since the SecureStream benchmark includes only a single query consisting of three operators.

\begin{table*}
\centering
\scriptsize
\caption{Query Recovery Success Rate (QRSR) in the first setting}
\label{tab:qrsr_first_setting}
\begin{tabular}{cccccclccclccc}
\hline
\multirow{2}{*}{\textbf{Query}} & \multirow{2}{*}{\textbf{Operators}} & \multirow{2}{*}{\textbf{\begin{tabular}[c]{@{}c@{}}No. \\Samples\end{tabular}}} & \multicolumn{3}{c}{\textbf{Random Forest}}                                                                                                                                           &                      & \multicolumn{3}{c}{\textbf{SVM}}                                                                                                                                                     &                      & \multicolumn{3}{c}{\textbf{XGBoost}}                                                                                                                                                     \\ \cline{4-6} \cline{8-10} \cline{12-14} 
                                &                                     &                                       & \textbf{\begin{tabular}[c]{@{}c@{}}Correctly\\ Predicted\end{tabular}} & \textbf{\begin{tabular}[c]{@{}c@{}}\% Correctly\\ Predicted\end{tabular}} & \textbf{QRSR}                   &                      & \textbf{\begin{tabular}[c]{@{}c@{}}Correctly\\ Predicted\end{tabular}} & \textbf{\begin{tabular}[c]{@{}c@{}}\% Correctly\\ Predicted\end{tabular}} & \textbf{QRSR}                   &                      & \textbf{\begin{tabular}[c]{@{}c@{}}Correctly\\ Predicted\end{tabular}} & \textbf{\begin{tabular}[c]{@{}c@{}}\% Correctly\\ Predicted\end{tabular}} & \textbf{QRSR}                   \\ \hline
\multicolumn{14}{c}{\textbf{CDF}}                                                                                                                                                                                                                                                                                                                                                                                                                                                                                                                                                                                                                                                                                                \\ \hline
Q1                              & Map                                 & 123                                   & 118                                                                    & 95.93                                                                     & \textbf{95.93}                  &                      & 108                                                                    & 87.80                                                                     & \textbf{87.80}                  &                      & 118                                                                    & 95.93                                                                     & \textbf{95.93}                  \\ \hline
\multirow{2}{*}{Q2}             & Filter                              & 152                                   & 135                                                                    & 88.82                                                                     & \multirow{2}{*}{\textbf{88.82}} &                      & 110                                                                    & 72.37                                                                     & \multirow{2}{*}{\textbf{70.63}} &                      & 123                                                                    & 80.92                                                                     & \multirow{2}{*}{\textbf{80.92}} \\
                                & Map                                 & 125                                   & 125                                                                    & 100                                                                       &                                 &                      & 122                                                                    & 97.6                                                                      &                                 &                      & 125                                                                    & 100                                                                       &                                 \\ \hline
\multirow{4}{*}{Q3}             & Filter 1                            & 186                                   & 185                                                                    & 99.46                                                                     & \multirow{4}{*}{\textbf{97.26}} &                      & 172                                                                    & 92.47                                                                     & \multirow{4}{*}{\textbf{49.81}} &                      & 185                                                                    & 99.46                                                                     & \multirow{4}{*}{\textbf{97.77}} \\
                                & Filter 2                            & 150                                   & 149                                                                    & 99.33                                                                     &                                 &                      & 142                                                                    & 94.67                                                                     &                                 &                      & 149                                                                    & 99.33                                                                     &                                 \\
                                & Join                                & 252                                   & 252                                                                    & 100                                                                       &                                 &                      & 148                                                                    & 58.73                                                                     &                                 &                      & 252                                                                    & 100                                                                       &                                 \\
                                & Map                                 & 192                                   & 189                                                                    & 98.44                                                                     &                                 &                      & 186                                                                    & 96.88                                                                     &                                 &                      & 190                                                                    & 98.96                                                                     &                                 \\ \hline
\multirow{4}{*}{Q4}             & Join                                & 253                                   & 253                                                                    & 100                                                                       & \multirow{4}{*}{\textbf{97.53}} &                      & 249                                                                    & 98.42                                                                     & \multirow{4}{*}{\textbf{90.00}} &                      & 251                                                                    & 99.21                                                                     & \multirow{4}{*}{\textbf{86.47}} \\
                                & Map                                 & 192                                   & 188                                                                    & 97.92                                                                     &                                 &                      & 178                                                                    & 92.71                                                                     &                                 &                      & 168                                                                    & 87.5                                                                      &                                 \\
                                & Max                                 & 256                                   & 256                                                                    & 100                                                                       &                                 &                      & 253                                                                    & 98.83                                                                     &                                 &                      & 256                                                                    & 100                                                                       &                                 \\
                                & Average                             & 509                                   & 507                                                                    & 99.61                                                                     &                                 &                      & 508                                                                    & 99.80                                                                     &                                 &                      & 507                                                                    & 99.61                                                                     &                                 \\ \hline
Q5                              & Count                               & 375                                   & 375                                                                    & 100                                                                       & \textbf{100}                    &                      & 372                                                                    & 99.2                                                                      & \textbf{99.2}                   &                      & 375                                                                    & 100                                                                       & \textbf{100}                    \\ \hline
\multirow{3}{*}{Q6}             & Filter                              & 161                                   & 136                                                                    & 84.47                                                                     & \multirow{3}{*}{\textbf{84.47}} &                      & 138                                                                    & 85.71                                                                     & \multirow{3}{*}{\textbf{84.59}} &                      & 121                                                                    & 75.16                                                                     & \multirow{3}{*}{\textbf{75.02}} \\
                                & Max                                 & 547                                   & 547                                                                    & 100                                                                       &                                 &                      & 547                                                                    & 100                                                                       &                                 &                      & 546                                                                    & 99.82                                                                     &                                 \\
                                & Average (P)                         & 536                                   & 536                                                                    & 100                                                                       &                                 &                      & 529                                                                    & 98.69                                                                     &                                 &                      & 536                                                                    & 100                                                                       &                                 \\ \hline
\multicolumn{14}{c}{\textbf{TS2Vec Embedding}}                                                                                                                                                                                                                                                                                                                                                                                                                                                                                                                                                                                                                                                                                   \\ \hline
Q1                              & Map                                 & 123                                   & 119                                                                    & 96.75                                                                     & \textbf{96.75}                  &                      & 113                                                                    & 91.87                                                                     & \textbf{91.87}                  &                      & 116                                                                    & 94.31                                                                     & \textbf{94.31}                  \\ \hline
\multirow{2}{*}{Q2}             & Filter                              & 152                                   & 121                                                                    & 79.61                                                                     & \multirow{2}{*}{\textbf{78.33}} &                      & 122                                                                    & 80.26                                                                     & \multirow{2}{*}{\textbf{76.41}} &                      & 120                                                                    & 78.94                                                                     & \multirow{2}{*}{\textbf{78.32}} \\
                                & Map                                 & 125                                   & 123                                                                    & 98.40                                                                     &                                 & \multicolumn{1}{c}{} & 119                                                                    & 95.20                                                                     &                                 & \multicolumn{1}{c}{} & 124                                                                    & 99.2                                                                      &                                 \\ \hline
\multirow{4}{*}{Q3}             & Filter 1                            & 186                                   & 136                                                                    & 73.12                                                                     & \multirow{4}{*}{\textbf{63.88}} &                      & 152                                                                    & 81.72                                                                     & \multirow{4}{*}{\textbf{66.28}} &                      & 143                                                                    & 76.88                                                                     & \multirow{4}{*}{\textbf{64.47}} \\
                                & Filter 2                            & 150                                   & 136                                                                    & 90.67                                                                     &                                 & \multicolumn{1}{c}{} & 138                                                                    & 92.00                                                                     &                                 & \multicolumn{1}{c}{} & 136                                                                    & 90.67                                                                     &                                 \\
                                & Join                                & 252                                   & 252                                                                    & 100                                                                       &                                 & \multicolumn{1}{c}{} & 248                                                                    & 98.41                                                                     &                                 & \multicolumn{1}{c}{} & 250                                                                    & 99.21                                                                     &                                 \\
                                & Map                                 & 192                                   & 185                                                                    & 96.35                                                                     &                                 & \multicolumn{1}{c}{} & 172                                                                    & 89.58                                                                     &                                 & \multicolumn{1}{c}{} & 179                                                                    & 93.23                                                                     &                                 \\ \hline
\multirow{4}{*}{Q4}             & Join                                & 253                                   & 253                                                                    & 100                                                                       & \multirow{4}{*}{\textbf{85.03}} &                      & 252                                                                    & 99.60                                                                     & \multirow{4}{*}{\textbf{80.69}} &                      & 253                                                                    & 100                                                                       & \multirow{4}{*}{\textbf{82.80}} \\
                                & Map                                 & 192                                   & 188                                                                    & 97.92                                                                     &                                 & \multicolumn{1}{c}{} & 178                                                                    & 92.71                                                                     &                                 & \multicolumn{1}{c}{} & 184                                                                    & 95.83                                                                     &                                 \\
                                & Max                                 & 256                                   & 256                                                                    & 100                                                                       &                                 & \multicolumn{1}{c}{} & 247                                                                    & 96.48                                                                     &                                 & \multicolumn{1}{c}{} & 253                                                                    & 98.828125                                                                 &                                 \\
                                & Average                             & 509                                   & 442                                                                    & 86.84                                                                     &                                 & \multicolumn{1}{c}{} & 461                                                                    & 90.57                                                                     &                                 & \multicolumn{1}{c}{} & 445                                                                    & 87.43                                                                     &                                 \\ \hline
Q5                              & Count                               & 375                                   & 374                                                                    & 99.73                                                                     & \textbf{99.73}                  &                      & 374                                                                    & 99.73                                                                     & \textbf{99.73}                  &                      & 374                                                                    & 99.73                                                                     & \textbf{99.73}                  \\ \hline
\multirow{3}{*}{Q6}             & Filter                              & 161                                   & 119                                                                    & 73.91                                                                     & \multirow{3}{*}{\textbf{73.51}} &                      & 132                                                                    & 81.99                                                                     & \multirow{3}{*}{\textbf{79.29}} &                      & 121                                                                    & 75.16                                                                     & \multirow{3}{*}{\textbf{74.04}} \\
                                & Max                                 & 547                                   & 545                                                                    & 99.63                                                                     &                                 &                      & 534                                                                    & 97.62                                                                     &                                 &                      & 543                                                                    & 99.27                                                                     &                                 \\
                                & Average (P)                         & 536                                   & 535                                                                    & 99.81                                                                     &                                 &                      & 531                                                                    & 99.07                                                                     &                                 &                      & 532                                                                    & 99.25                                                                     &                                 \\ \hline
\end{tabular}
\end{table*}

Table \ref{tab:qrsr_first_setting} shows the results under the first setting. Using CDF, all models achieved a QRSR of at least 80\% across all queries, except for the XGBoost model on Query 6 (Q6). Among the tested models, Random Forest consistently outperformed others, demonstrating dominant performance in nearly all cases. Notably, using CDF results in a high QRSR even in complex queries such as Q3 and Q4, both of which involve multiple operators. TS2Vec leads to poorer performance. In particular, three out of six queries have QRSR values below 80\%. These findings diverge from the results reported earlier in Table \ref{tab:performances_single_benchmark}, where TS2Vec outperforms CDF.

\begin{table*}
\centering
\scriptsize
\caption{Query Recovery Success Rate (QRSR) in the second setting}
\label{tab:qrsr_second_setting}
\begin{tabular}{cccccccccccccc}
\hline
\multirow{2}{*}{\textbf{Query}} & \multirow{2}{*}{\textbf{Operators}} & \multirow{2}{*}{\textbf{\begin{tabular}[c]{@{}c@{}}No. \\Samples\end{tabular}}} & \multicolumn{3}{c}{\textbf{Random Forest}}                                                                                                                                             & \textbf{} & \multicolumn{3}{c}{\textbf{SVM}}                                                                                                                                                       & \textbf{} & \multicolumn{3}{c}{\textbf{XGBoost}}                                                                                                                                                       \\ \cline{4-6} \cline{8-10} \cline{12-14} 
                                &                                     &                                 & \textbf{\begin{tabular}[c]{@{}c@{}}Correctly\\ Classified\end{tabular}} & \textbf{\begin{tabular}[c]{@{}c@{}}\% Correctly\\ Classified\end{tabular}} & \textbf{QRSR}                   & \textbf{} & \textbf{\begin{tabular}[c]{@{}c@{}}Correctly\\ Classified\end{tabular}} & \textbf{\begin{tabular}[c]{@{}c@{}}\% Correctly\\ Classified\end{tabular}} & \textbf{QRSR}                   & \textbf{} & \textbf{\begin{tabular}[c]{@{}c@{}}Correctly\\ Classified\end{tabular}} & \textbf{\begin{tabular}[c]{@{}c@{}}\% Correctly\\ Classified\end{tabular}} & \textbf{QRSR}                   \\ \hline
\multicolumn{14}{c}{CDF}                                                                                                                                                                                                                                                                                                                                                                                                                                                                                                                                                                                                                                                                                   \\ \hline
Q1                              & Map                                 & 245                             & 144                                                                     & 58.78                                                                      & \textbf{58.78}                  &           & 211                                                                     & 86.12                                                                      & \textbf{86.12}                  &           & 118                                                                     & 48.16                                                                      & \textbf{48.16}                  \\ \hline
\multirow{2}{*}{Q2}             & Filter                              & 304                             & 211                                                                     & 69.41                                                                      & \multirow{2}{*}{\textbf{69.41}} &           & 167                                                                     & 54.93                                                                      & \multirow{2}{*}{\textbf{53.61}} &           & 146                                                                     & 48.03                                                                      & \multirow{2}{*}{\textbf{48.03}} \\
                                & Map                                 & 249                             & 249                                                                     & 100                                                                        &                                 &           & 243                                                                     & 97.59                                                                      &                                 &           & 249                                                                     & 100                                                                        &                                 \\ \hline
\multirow{4}{*}{Q3}             & Filter 1                            & 372                             & 347                                                                     & 93.28                                                                      & \multirow{4}{*}{\textbf{0.09}}  &           & 234                                                                     & 62.90                                                                      & \multirow{4}{*}{\textbf{0.99}}  &           & 259                                                                     & 69.62                                                                      & \multirow{4}{*}{\textbf{0.29}}  \\
                                & Filter 2                            & 300                             & 207                                                                     & 69.00                                                                      &                                 &           & 131                                                                     & 43.67                                                                      &                                 &           & 103                                                                     & 34.33                                                                      &                                 \\
                                & Join                                & 504                             & 10                                                                      & 1.98                                                                       &                                 &           & 38                                                                      & 7.54                                                                       &                                 &           & 15                                                                      & 2.98                                                                       &                                 \\
                                & Map                                 & 384                             & 28                                                                      & 7.29                                                                       &                                 &           & 183                                                                     & 47.66                                                                      &                                 &           & 158                                                                     & 41.15                                                                      &                                 \\ \hline
\multirow{3}{*}{Q4}             & Join                                & 505                             & 251                                                                     & 49.70                                                                      & \multirow{3}{*}{\textbf{4.24}}  &           & 217                                                                     & 42.97                                                                      & \multirow{3}{*}{\textbf{22.14}} &           & 10                                                                      & 1.98                                                                       & \multirow{3}{*}{\textbf{0}}     \\
                                & Map                                 & 384                             & 40                                                                      & 10.42                                                                      &                                 &           & 287                                                                     & 74.74                                                                      &                                 &           & 0                                                                       & 0                                                                          &                                 \\
                                & Max                                 & 512                             & 419                                                                     & 81.84                                                                      &                                 &           & 353                                                                     & 68.95                                                                      &                                 &           & 494                                                                     & 96.48                                                                      &                                 \\ \hline
\multirow{2}{*}{Q6}             & Filter                              & 321                             & 142                                                                     & 44.24                                                                      & \multirow{2}{*}{\textbf{12.09}} &           & 139                                                                     & 43.30                                                                      & \multirow{2}{*}{\textbf{36.97}} &           & 7                                                                       & 2.18                                                                       & \multirow{2}{*}{\textbf{2.08}}  \\
                                & Max                                 & 1094                            & 299                                                                     & 27.33                                                                      &                                 &           & 934                                                                     & 85.37                                                                      &                                 &           & 1045                                                                    & 95.52                                                                      &                                 \\ \hline
\multicolumn{14}{c}{\textbf{TS2Vec}}                                                                                                                                                                                                                                                                                                                                                                                                                                                                                                                                                                                                                                                                       \\ \hline
Q1                              & Map                                 & 245                             & 233                                                                     & 95.10                                                                      & \textbf{95.10}                  &           & 200                                                                     & 81.63                                                                      & \textbf{81.63}                  &           & 221                                                                     & 90.20                                                                      & \textbf{90.20}                  \\ \hline
\multirow{2}{*}{Q2}             & Filter                              & 304                             & 231                                                                     & 75.99                                                                      & \multirow{2}{*}{\textbf{75.68}} &           & 236                                                                     & 77.63                                                                      & \multirow{2}{*}{\textbf{73.89}} &           & 236                                                                     & 77.63                                                                      & \multirow{2}{*}{\textbf{77.01}} \\
                                & Map                                 & 249                             & 248                                                                     & 99.60                                                                      &                                 &           & 237                                                                     & 95.18                                                                      &                                 &           & 247                                                                     & 99.20                                                                      &                                 \\ \hline
\multirow{4}{*}{Q3}             & Filter 1                            & 372                             & 250                                                                     & 67.20                                                                      & \multirow{4}{*}{\textbf{8.17}}  &           & 252                                                                     & 67.74                                                                      & \multirow{4}{*}{\textbf{8.40}}  &           & 251                                                                     & 67.47                                                                      & \multirow{4}{*}{\textbf{7.86}}  \\
                                & Filter 2                            & 300                             & 266                                                                     & 88.67                                                                      &                                 &           & 255                                                                     & 85.00                                                                      &                                 &           & 272                                                                     & 90.67                                                                      &                                 \\
                                & Join                                & 504                             & 78                                                                      & 15.48                                                                      &                                 &           & 96                                                                      & 19.05                                                                      &                                 &           & 80                                                                      & 15.87                                                                      &                                 \\
                                & Map                                 & 384                             & 340                                                                     & 88.54                                                                      &                                 &           & 294                                                                     & 76.56                                                                      &                                 &           & 311                                                                     & 80.99                                                                      &                                 \\ \hline
\multirow{3}{*}{Q4}             & Join                                & 505                             & 504                                                                     & 99.80                                                                      & \multirow{3}{*}{\textbf{87.34}} &           & 489                                                                     & 96.83                                                                      & \multirow{3}{*}{\textbf{82.90}} &           & 503                                                                     & 99.60                                                                      & \multirow{3}{*}{\textbf{83.66}} \\
                                & Map                                 & 384                             & 349                                                                     & 90.89                                                                      &                                 &           & 338                                                                     & 88.02                                                                      &                                 &           & 327                                                                     & 85.16                                                                      &                                 \\
                                & Max                                 & 512                             & 493                                                                     & 96.29                                                                      &                                 &           & 498                                                                     & 97.27                                                                      &                                 &           & 505                                                                     & 98.63                                                                      &                                 \\ \hline
\multirow{2}{*}{Q6}             & Filter                              & 321                             & 162                                                                     & 50.47                                                                      & \multirow{2}{*}{\textbf{32.71}} &           & 155                                                                     & 48.29                                                                      & \multirow{2}{*}{\textbf{37.96}} &           & 174                                                                     & 54.21                                                                      & \multirow{2}{*}{\textbf{35.43}} \\
                                & Max                                 & 1094                            & 709                                                                     & 64.81                                                                      &                                 &           & 860                                                                     & 78.61                                                                      &                                 &           & 715                                                                     & 65.36                                                                      &                                 \\ \hline
\end{tabular}
\end{table*}

Table \ref{tab:qrsr_second_setting} reports the results under the second experimental setting. Query 5 (Q5) is excluded because it only consists of the Count operator, which does not appear in any other query. Similarly, the Average (Partition) operator of Q6 is excluded due to its unique appearance. The results show a sharp drop in performance for CDF across most queries. The performance of Q3 and Q4, the two most structurally complex queries in the benchmark, is particularly poor. For Q3, classification accuracy is so low that no model achieves more than 1\% accuracy, making query recovery impossible. Similarly, in Q4, the XGBoost model fails to detect any instances of the Map operator, resulting in a QRSR of zero. Overall, except for Q2, where Random Forest performs the best, SVM consistently and significantly surpasses other models.

The results for TS2Vec are relatively better. In three of the five evaluated queries, all models achieve QRSR scores above 70\%, including Q4. However, the performance of all models decreases significantly in Q3 and Q6, indicating that the structural complexity and operator diversity of these queries remain a fundamental challenge. Unlike CDF, no single model under TS2Vec clearly dominates the others across the majority of queries.

It can be seen that the models perform better under the first setting than the second. This is expected because, in the first setting, models are trained on a subset of the same operators they later predict, while in the second setting, they must generalize to unseen queries, which is more challenging. The second setting represents a more realistic attack scenario, in which the attacker first profiles operators from their own queries or from widely available benchmark queries, then predicts an unseen query. The results in Table \ref{tab:qrsr_second_setting} suggest that, under such conditions, an attacker can practically recover a large number of target queries.


\subsubsection{Predicting operator parameters}
We present the performance of operator parameter inference models in Table \ref{tab:window_size_pred_performance} for window sizes. Additional results for sliding steps are presented in Appendix \ref{app:evalution}. We report Mean Squared Error (MSE) (scaled by $\times10^5$) and R² score for each model across both CDF-based and TS2Vec-based representations. Two parameters are evaluated: window size and sliding step.

For window size prediction, the TS2Vec-based Random Forest model outperforms other configurations in most cases, achieving a $R^2$ score exceeding 95\% for the Join, Average (Sliding), and Count operators. Even for the Average (Partitioned) operator, where performance drops slightly, the model still attains a $R^2$ score of 85\%, the highest among all models. The only exception is the Max operator, where this model performs poorly and ranks third, while the TS2Vec-based SVM achieves the best performance with a $R^2$ score of 85\%. Overall, other models—both CDF-based and TS2Vec-based—demonstrate comparable but generally lower performance in this task.

These results suggest that Random Forest is the most suitable model for parameter inference tasks. However, neither representation method—CDF nor TS2Vec—performs optimally across all cases. This limitation is particularly important given that the experiment assumes the operator type is known a priori, an ideal scenario from the attacker's perspective. In practical settings, operator misclassification may occur, potentially leading to erroneous parameter inference. Addressing this limitation and improving robustness under such uncertainty is a promising direction for future work.

\begin{table*}
\centering
\scriptsize
\caption{Performance of window size prediction models on NEXMark}
\label{tab:window_size_pred_performance}
\begin{tabular}{lcccccccccc}
\hline
\multicolumn{1}{c}{\multirow{2}{*}{\textbf{\begin{tabular}[c]{@{}c@{}}Window Size\\ Prediction Models\end{tabular}}}} & \multicolumn{2}{c}{\textbf{Join}}         & \multicolumn{2}{c}{\textbf{Max}}          & \multicolumn{2}{c}{\textbf{Average}}  & \multicolumn{2}{c}{\textbf{Avg (Partition)}} & \multicolumn{2}{c}{\textbf{Count}}        \\ \cline{2-11} 
\multicolumn{1}{c}{}                                                                                                  & \textbf{MSE ($\times10^5$)} & \textbf{R2} & \textbf{MSE ($\times10^5$)} & \textbf{R2} & \textbf{MSE ($\times10^5$)} & \textbf{R2} & \textbf{MSE ($\times10^5$)}  & \textbf{R2}  & \textbf{MSE ($\times10^5$)} & \textbf{R2} \\ \hline
\multicolumn{11}{c}{\textbf{CDF}}                                                                                                                                                                                                                                                                                                                   \\ \hline
Random Forest                                                                                                         & 0.0094                      & 0.9549      & 74.1960                     & -2.6152     & 10.3621                     & 0.9723      & 1.1852                       & 0.8245       & 0.0005                      & 0.9746      \\
SVM                                                                                                                   & 13.4381                     & -63.4864    & 62.6543                     & -2.0528     & 32.0333                     & 0.9144      & 2.0173                       & 0.7013       & 0.0005                      & 0.9721      \\
XGBoost                                                                                                               & 0.0155                      & 0.9255      & 72.2303                     & -2.5194     & 0.9517                      & 0.9746      & 1.2752                       & 0.8112       & 0.0004                      & 0.9771      \\ \hline
\multicolumn{11}{c}{\textbf{TS2Vec Embedding}}                                                                                                                                                                                                                                                                                                      \\ \hline
Random Forest                                                                                                         & \textbf{0.0075}                      & \textbf{0.9640}      & 12.8485                     & 0.3740      & \textbf{0.1787}                      & \textbf{0.9950}      & \textbf{1.0144}                       & \textbf{0.8552}       & \textbf{0.0003}                      & \textbf{0.9847}      \\
SVM                                                                                                                   & 1.4759                      & 0.9277      & \textbf{2.5590}                      & \textbf{0.8537}      & 2.4879                      & 0.9358      & 1.1225                       & 0.8396       & 0.0007                      & 0.9585      \\
XGBoost                                                                                                               & 0.9232                      & -3.4300     & 9.4244                      & 0.5408      & 2.0530                      & 0.9421      & 1.7133                       & 0.7554       & 0.0198                      & -0.0951     \\ \hline
\end{tabular}
\end{table*}

\section{Discussions}
\label{sec:discussion}

\noindent \textbf{Insecure privacy-preserving implementation}.
From the experimental results in \S\ref{subsec:results}, we observe that the success of the attack depends on mismatched execution latencies among operations when deployed across individual enclaves.
Therefore, even naive adaptations of existing privacy-preserving techniques, such as HE and secure multiparty computation, are insufficient to mitigate these attacks.
For example, one might consider implementing secret sharing or functional encryption \cite{Xing25,Boyle25,Gilboa14,Boyle16} across multiple enclaves to ensure data confidentiality and integrity.
However, if both parties jointly execute the same operation, e.g.,  \textit{Max} operation as in Figure \ref{fig:cdf_of_each_operators} with identical computational complexity, an attacker can still distinguish this from the secure computation of other operations, e.g., from the \textit{Map} operation.
Therefore, the mitigation should be obfuscating the query operations to prevent the attacks.

\vspace{0.1cm}
\noindent \textbf{Obfuscation}.
The first simple-yet solution is to pad dummy operations to ensure indistinguishable execution latency among operations. 
For example, the enclave can leverage both the oblivious data structure, i.e., ORAM and its variations~\cite{Elaine20}, and existing instruction sets in Intel SGX~\cite{Ohrimenko16,Sasy17,Rane15} to build new oblivious primitives for comparisons, assignment, sort, and selection. 
These primitives could be combined together to protect the query operations.

Another mitigation is by preventing the DAG from being known or isolating operators from queries.
For example, the cloud provider can employ new oblivious scheduling policies to ensure a certain number of operators must be batched and executed per enclave to ensure uniform computational complexity distribution across enclaves.
Alternatively, to prevent the operation profiling, one can make custom code by merging or de-modularizing operators, e.g., Map and Filter logic, to avoid distinguishability.
%

We stress that all the above mitigation approaches typically require a significant effort and dedication to review the implementation of streaming applications.

\vspace{0.1cm}
\noindent \textbf{Attack limitations}.
There are certain factors omitted during the attack experiments.
At first, our proposed attacks are experimented with in isolated Intel SGX enclaves without other background processes.
In addition, the attacks have not been experimented with and evaluated on commercial TEE-enabled clouds for full-scale DSP systems.
Moreover, the attacks also have not been evaluated in the second generation of Intel SGX and other trusted platforms like ARM Zone.
Therefore, we believe these factors may add noise to the execution latency due to both hardware and context switching, which certainly reduces the attack prediction accuracy.
A promising complement solution is to further investigate the capabilities of ML and recent AI foundation models in this attack and mitigation directions \cite{Umar25,Wu25}.

\section{Related Work}
\label{sec:related}
\subsubsection*{Distributed stream processing systems}
Apache Storm~\cite{toshniwal2014storm} is among the first distributed stream processing systems. It adopts the execution model in 
which the query is represented as a DAG of operators, and data flows from one operator to the next. Subsequent systems 
follow the same model and add more features to the original Storm design. Spark Streaming \cite{zaharia2013discretized}, 
Flink \cite{carbone2015apache} and Samza \cite{noghabi2017samza} support both batch and stream processing in a single framework and include
many stateful operators such as joins and aggregates. These systems handle operator failures, which can affect query 
correctness, by replicating the operators, checkpointing \cite{murray2013naiad}, or a combination thereof \cite{lin2016streamscope}. Other works focus on improving the performance of DAG 
execution by detecting bottlenecks via backpressure \cite{kulkarni2015twitter}, or by dynamically replicating the
operators to adapt to workload changes \cite{castro2013integrating,wu2015chronostream}. Our work demonstrates attacks 
against a generic DSP system, which can be applied to systems other than Flink.  

\subsubsection*{Outsourced data processing}
Designing an outsourced database running on an untrusted cloud requires making trade-offs between security, 
functionality, and performance. Cryptographic solutions such as CryptDB \cite{popa2011cryptdb} and 
Arx \cite{poddar2016arx} protect data confidentiality using cryptographic techniques, namely, using homomorphic and
functional encryption schemes. However, they support limited computation over the data, and have been shown to leak data 
in practical settings \cite{naveed2015inference}. Secure databases built on top of trusted hardware include 
Cipherbase~\cite{arasu2013orthogonal} and TrustedDB~\cite{bajaj2011trusteddb} that use older hardware.                      
More recent systems based on Intel SGX, such as ObliDB \cite{oblidb}, Opaque \cite{zheng2017opaque}, Obaldi \cite{crooks2018obladi}, focus on obliviousness to prevent leakage through memory access patterns \cite{priebe2018enclavedb}. They achieve high performance and more functionalities than purely cryptographic solutions, but their security assumptions are stronger, i.e. the hardware is secure. Furthermore, they incur non-trivial
overhead due to obliviousness.

Our work focuses data stream processing systems --- one type of database that handles continuous, infinite streams of 
data. In these systems, query confidentiality, which is overlooked in the other works targeting relational databsaes, is 
as important as data confidentiality. StreamBox-TZ~\cite{234964} and SecureStream~\cite{DBLP:journals/corr/abs-1805-01752} are the two most 
relevant streaming systems using secure hardware, but they also do not provide query confidentiality. 

\subsubsection*{Side channels}
Trusted hardware, in particular Intel SGX, has been a popular target for side channel attacks. The attacker gains
knowledge of the code and data running inside the enclaves via both hardware and software side channels. Since SGX
assumes a strong threat model, the attacker can manipulate page faults~\cite{van2017telling}, timer                         interrupt~\cite{van2017sgx}, and exception~\cite{cui2021smashex} to leak content from inside the enclave. Despite the           closed design, recent works have uncovered multiple hardware, micro-architectural side channels, including hardware              
cache~\cite{brasser2017software, van2021cacheout}, transient execution~\cite{van2018foreshadow}, and branch
prediction~\cite{lee2017inferring}. A comprehensive list of attacks against SGX can be found in~\cite{van2024sok}. Our 
work demonstrates another side channel specific to distributed stream processing systems. It is simpler to execute than 
other state-of-the-art channels, and we demonstrate that it is highly effective at leaking the application queries. How 
our attack can be improved with the other side channels is left as future work.
\section{Conclusions}
\label{sec:conclusions}
In this paper, we demonstrated a side channel attack that leaks encrypted queries from a streaming application. The query runs as a DAG of operators on a distributed stream processing system, which executes the operators inside SGX enclaves. The attacker profiles the execution timing of individual operators based on its synthetic data. It then trains machine learning models that predict the operator's identity and internal parameters. Our experiments show that the attack is effective, with success rates up to 92\%. We then discussed several system-level mitigation approaches that can defend against the attack, thereby making the first steps toward securing modern streaming applications.


\bibliographystyle{IEEEtran}
\bibliography{main_full_version}


\begin{appendices}

\section{Benchmark}
\label{app:benchmark}
\subsubsection*{SecureStream}
The pipeline consists of three operators. The \textbf{\textit{Map}} operator transforms each raw CSV line into a structured object containing only relevant fields (e.g., carrier ID and delay time), discarding unnecessary information. The \textbf{\textit{Filter}} operator evaluates each structured event and retains only those corresponding to delayed flights (i.e., delay time $>0$). The \textbf{\textit{Reduce}} operator computes aggregate statistics—specifically, the average delay duration—using a windowed processing mechanism.

The first two operators, Map and Filter, are stateless: they operate on a per-event basis and do not maintain any internal state between inputs. In contrast, the Reduce operator is stateful, as it processes a window of multiple events and must retain internal state (e.g., cumulative delay and count) to compute aggregate results. This distinction is critical, as the presence or absence of a state affects the processing time characteristics exploited in our attack.

\subsubsection*{NexMark}

In summary, our implementation of the NEXMark benchmark evaluates seven distinct stream processing operators (see Table~\ref{tab:NEXMark_summary}). Notably, the Average operator appears in both Q4 and Q6, but with different windowing mechanisms: Q4 employs an Average operator over all data events, whereas Q6 uses a Partitioning, or Keying, average mechanism. Given that different aggregation strategies significantly affect operator behavior and processing characteristics, we treat the same operation with different aggregation mechanisms as separate operator types in our analysis.

Unlike the SecureStream benchmark, which uses a pre-existing real-world dataset, the NEXMark dataset must be synthetically generated. While the original authors provide a Java-based XML generator, we opted to use the \textit{NEXMark Event Generator} \cite{nexmark_data_generator}, a Rust-based tool that offers a more convenient command-line interface and produces output in a structured JSON format. This tool allowed us to generate a substantial volume of data: over 20,000 Person events, 60,000 Auction events, and 920,000 Bid events, which were used as input for the selected queries.

\section{Evaluation}
\label{app:evalution}

\subsection{Hyperparameters}
Table~\ref{tab:model_hyperparameters} lists the parameters of the models. 

\begin{table}[htb]
\centering
\caption{Model Hyperparameters}
\label{tab:model_hyperparameters}
\begin{tabular}{lll}
\hline
\multicolumn{1}{c}{\textbf{Models}} & \textbf{Hyperparameters} & \textbf{Values}                  \\ \hline
\multirow{4}{*}{Random Forest}      & no. of estimators        & 100, 150, 200, 250, 300          \\
                                    & max depth                & no limit, 10, 20                 \\
                                    & max features             & sqrt, log2                       \\
                                    & booststrap               & true, false                      \\ \hline
\multirow{3}{*}{SVM}                & C                        & 0.1, 1, 10, 100                  \\
                                    & gamma                    & scale, auto, 0.001, 0.01, 0.1, 1 \\
                                    & kernel                   & linear, rbf, poly                \\ \hline
\multirow{3}{*}{XGBoost}            & no. of estimators        & 100, 150, 200, 250, 300          \\
                                    & max depth                & 3, 6, 10                         \\
                                    & gamma                    & 0, 1, 5                          \\ \hline
\end{tabular}
\end{table}

\begin{table*}
\centering
\caption{Summarization of deployed queries in NEXMark}
\label{tab:NEXMark_summary}
\begin{tabular}{lll}
\hline
\multicolumn{1}{c}{\textbf{No}} & \multicolumn{1}{c}{\textbf{Summary}}                                                                                                                             & \multicolumn{1}{c}{\textbf{Operators}} \\ \hline
Q1                              & Convert the value of each Bid from dollars to euros                                                                                                              & Map                                    \\
Q2                              & Find Bids belong to specific Auctions, then return those Auctions and Bids' price                                                                                & Filter, Map                            \\
Q3                              & \begin{tabular}[c]{@{}l@{}}Find People who are selling items of a specific category and their Auctions,\\ then return specific People's information\end{tabular} & Filter, Join, Map                      \\
Q4                              & Calculate the average prices of winning Bids of all Auctions                                                                                                     & Join, Max, Average                     \\
Q5                              & Count Auctions that have the most Bids                                                                                                                           & Count                                  \\
Q6                              & Calculate the average price per Person/seller                                                                                                                    & Join, Filter, Max, Average (Partition) \\ \hline
\end{tabular}
\end{table*}

\subsection{Confusion Matrix} 
\begin{figure}
    \centering
    \begin{subfigure}[t]{0.4\textwidth}
        \includegraphics[width=\linewidth]{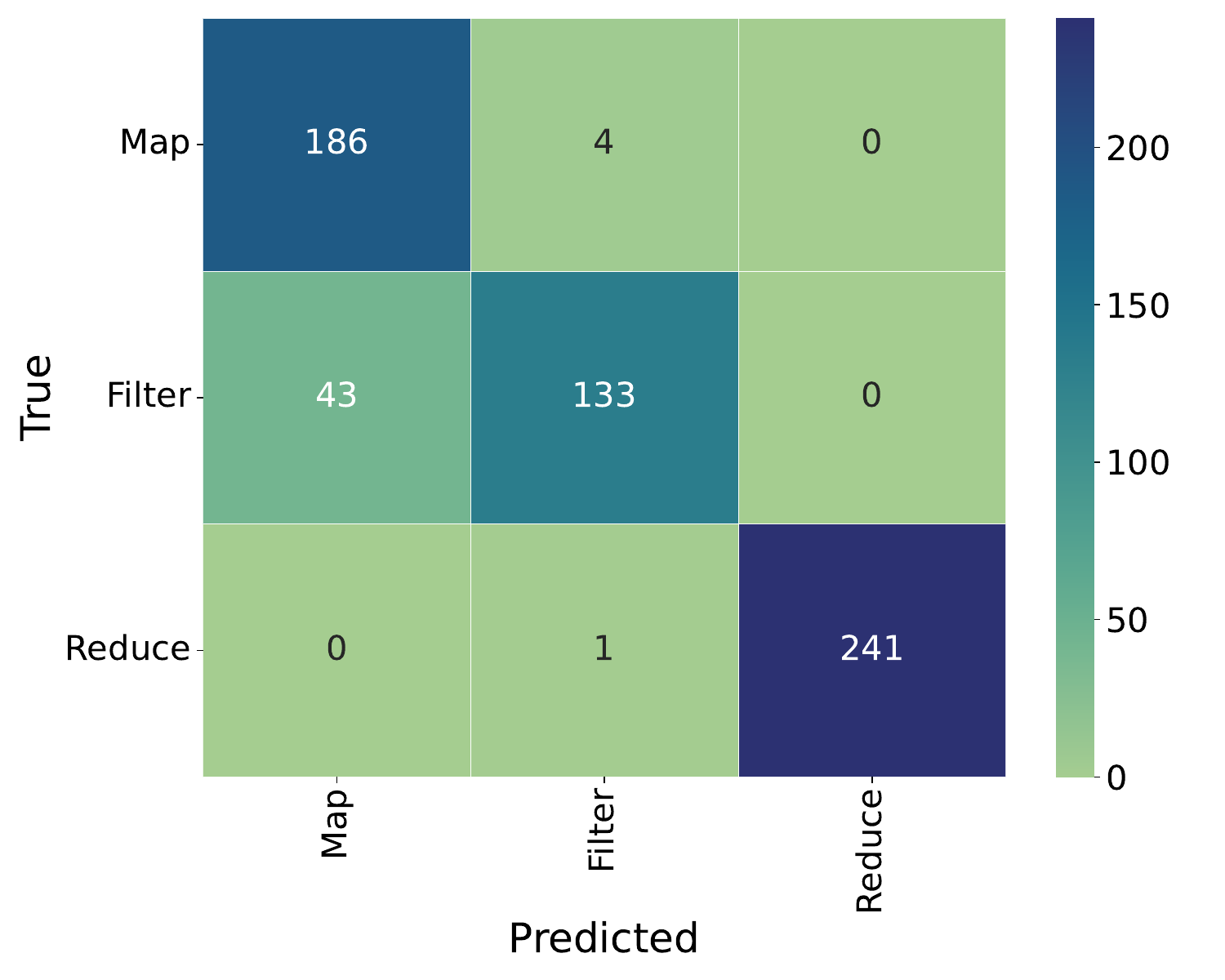}
        \caption{SecureStream}
    \end{subfigure}%

    \vspace{1mm}
    \begin{subfigure}[t]{0.45\textwidth}
        \includegraphics[width=\linewidth]{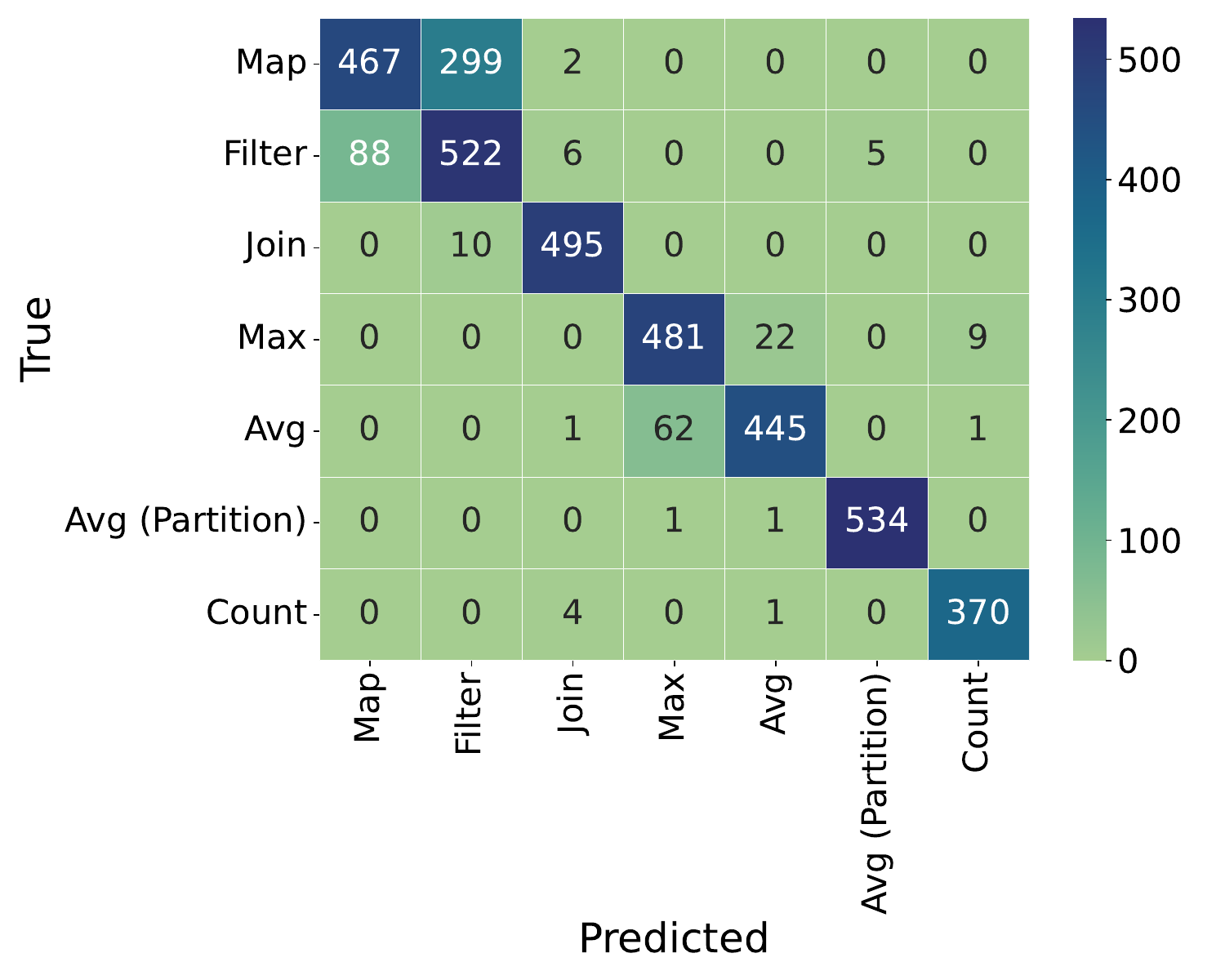}
        \caption{NEXMark}
    \end{subfigure}
    \caption{Confusion matrix} 
    \label{fig:confusion_matrix}
\end{figure}

To gain deeper insights into model performance, we examine the confusion matrices of the best-performing TS2Vec-based models for both benchmarks, which are presented in Figure \ref{fig:confusion_matrix}.

In the SecureStream benchmark, the model exhibits difficulty in distinguishing between the Map and Filter operators, which was anticipated from the t-SNE visualization. Specifically, 32\% of Filter samples are misclassified as Map. This confusion likely arises from the similarity in their stateless nature and processing characteristics. In the NEXMark benchmark, similar challenges are observed. The model misclassifies 64\% of Map samples and 17\% of Filter samples as the other class. Additionally, the model struggles to differentiate between the Max and Average operators. This may be attributed to their shared characteristics as aggregation functions operating over sliding windows. Interestingly, the Count operator—despite also being an aggregation function with the same windowing mechanism—is classified more accurately. This could be because it performs simpler logic (i.e., incrementing a count) without the need for value comparisons or numerical calculations, making its timing behavior more consistent and distinguishable.

\begin{table*}
\centering
\scriptsize
\caption{Performances of Sliding Step Prediction Models on NEXMark}
\label{tab:sliding_step_pred_perfomance}
\begin{tabular}{lcccccccccc}
\hline
\multicolumn{1}{c}{\multirow{2}{*}{\textbf{\begin{tabular}[c]{@{}c@{}}Sliding Step\\ Prediction Models\end{tabular}}}} & \multicolumn{2}{c}{\textbf{Join}} & \multicolumn{2}{c}{\textbf{Max}} & \multicolumn{2}{c}{\textbf{Average}} & \multicolumn{2}{c}{\textbf{Avg (Partition)}} & \multicolumn{2}{c}{\textbf{Count}} \\ \cline{2-11} 
\multicolumn{1}{c}{}                                                                                                   & \textbf{MSE ($\times10^5$)}     & \textbf{R2}    & \textbf{MSE ($\times10^5$)}    & \textbf{R2}    & \textbf{MSE ($\times10^5$)}        & \textbf{R2}        & \textbf{MSE ($\times10^5$)}          & \textbf{R2}         & \textbf{MSE ($\times10^5$)}     & \textbf{R2}     \\ \hline
\multicolumn{11}{c}{\textbf{CDF}}                                                                                                                                                                                                                                                                                           \\ \hline
Random Forest                                                                                                          & \textbf{0.0023}           & \textbf{0.7223}         & \textbf{0.0186}          & \textbf{0.9979}         & \textbf{0.0355}              & \textbf{0.9978}             & \textbf{0.0387}                & \textbf{0.8356}              & \textbf{0.0000}           & \textbf{0.9997}          \\
SVM                                                                                                                    & 0.3841           & -45.4959       & 6.5675          & 0.2528         & 0.6475              & 0.9599             & 0.1015                & 0.5693              & 0.00007          & 0.9911          \\
XGBoost                                                                                                                & 0.0065           & 0.2184         & 0.6448          & 0.9266         & 0.0560              & 0.9965             & 0.0398                & 0.8309              & 0.00008          & 0.9900          \\ \hline
\multicolumn{11}{c}{\textbf{TS2Vec Embedding}}                                                                                                                                                                                                                                                                            \\ \hline
Random Forest                                                                                                          & 1.1531           & -138.5797      & 3.5716          & 0.5936         & 7.9254              & 0.5261             & 0.0755                & 0.6550              & 0.0002           & 0.9737          \\
SVM                                                                                                                    & 0.9418           & -112.9975      & 5.5350          & 0.3703         & 12.7500             & 0.2376             & 0.0978                & 0.5529              & 0.0030           & 0.6480          \\
XGBoost                                                                                                                & 2.2325           & -269.2285      & 7.9687          & 0.0934         & 13.5072             & 0.1924             & 0.1155                & 0.4717              & 0.0072           & 0.1702          \\ \hline
\end{tabular}
\end{table*}

\subsection{Operator Parameter Prediction}
Table \ref{tab:sliding_step_pred_perfomance} shows the results for predicting sliding steps. For sliding step prediction, CDF-based models yield better results overall. In particular, the CDF-based Random Forest consistently outperforms other models across all operators, achieving $R^2$ scores above 99\% for Max, Average (Sliding), and Count. Its performance only declines for the Join operator, where the $R^2$ score drops to 72\%, though this still surpasses all other models. Notably, CDF-based XGBoost performs closely behind, typically within 0.5\% of the Random Forest model in 3/5 cases.

\end{appendices}

\end{document}